\documentclass[%
 aip,
 amsmath,amssymb,
 reprint,%
]{revtex4-2}

\usepackage{graphicx}% Include figure files
\usepackage{dcolumn}% Align table columns on decimal point
\usepackage{bm}% bold math
\usepackage[utf8]{inputenc}
\usepackage[T1]{fontenc}
\usepackage{mathptmx}
\usepackage{etoolbox}
\usepackage{siunitx}
\usepackage{color}
\usepackage{braket}
\usepackage[version=3]{mhchem}
\usepackage{multirow} % preambleに追加
\usepackage{booktabs,tabularx,array}
\usepackage{adjustbox}
\usepackage{makecell}
\usepackage{rotating}
\usepackage{makecell}
\usepackage{mhchem}

\makeatletter
\def\@email#1#2{%
 \endgroup
 \patchcmd{\titleblock@produce}
  {\frontmatter@RRAPformat}
  {\frontmatter@RRAPformat{\produce@RRAP{*#1\href{mailto:#2}{#2}}}\frontmatter@RRAPformat}
  {}{}
}%
\makeatother
             
\begin{document}

\preprint{AIP/123-QED}

\title[]{Engineering nanodiamonds for quantum sensing:\\ material constraints at the nanoscale}
% Force line breaks with \\

\author{Ashutosh Rathi}
\affiliation{ 
Department of Chemistry, Graduate School of Environmental, Life, Natural Science and Technology, 
Okayama University, Okayama 700-8530, Japan
}%

\author{Keisuke Oshimi}
\affiliation{
Department of Chemistry, Graduate School of Environmental, Life, Natural Science and Technology, 
Okayama University, Okayama 700-8530, Japan
}%
\affiliation{
Department of Physics, Humboldt-Universit\"{a}t zu Berlin, Berlin 12489, Germany
}%

\author{Kento Sasaki}
\author{Kensuke Kobayashi}
\affiliation{Department of Physics, The University of Tokyo, Bunkyo, Tokyo 113-0033, Japan}%

\author{Yutaka Shikano}
\affiliation{ 
Institute of Systems and Information Engineering, University of Tsukuba, Tsukuba, Ibaraki 305-8573, Japan
}%
\affiliation{ 
Center for Artificial Intelligence Research, University of Tsukuba, Tsukuba, Ibaraki 305-8577, Japan
}%
\affiliation{ 
Institute for Quantum Studies, Chapman University, 1 University Dr., Orange, California 92866, United States
}%

\author{Oliver Benson}
\affiliation{
Department of Physics, Humboldt-Universit\"{a}t zu Berlin, Berlin 12489, Germany
}%

\author{Tim Schr\"{o}der}
\affiliation{
Department of Physics, Humboldt-Universit\"{a}t zu Berlin, Berlin 12489, Germany
}%
\affiliation{
Ferdinand-Braun-Institut (FBH), Berlin 12489, Germany 
}%

\author{Shery L.Y. Chang}
\affiliation{
School of Materials Science and Engineering, UNSW Sydney, Sydney 2052, Australia
}%
\affiliation{
Electron Microscope Unit, Mark Wainwright Analytical Centre, University of New South Wales, Sydney 2052, Australia
}%

\author{Masazumi Fujiwara}
  \email{masazumi@okayama-u.ac.jp}
\affiliation{ 
Department of Chemistry, Graduate School of Environmental, Life, Natural Science and Technology, 
Okayama University, Okayama 700-8530, Japan
}%


\begin{abstract}

Optically addressable solid-state spin defects have emerged as powerful multimodal quantum sensors, with nitrogen--vacancy (NV) centers in bulk diamond providing benchmark quantum control and sensitivity under ambient conditions. Embedding such defects in nanodiamonds (NDs) extends these capabilities to mobile probes capable of accessing complex biological and nanoscale environments. Reduced dimensions, however, introduce constraints beyond volumetric spin impurities, notably enhanced lattice strain and surface-induced noise sources, which shorten NV spin relaxation times ($T_1$ and $T_2$) and destabilize the NV charge state, as well as resulting in pronounced particle-to-particle variability in NDs typically produced by top-down approaches. These effects complicate both sensing performance and the quantitative interpretation of multimodal signals in realistic environments.  This article provides a structured perspective on the physical mechanisms by which material properties constrain NV behavior in NDs, together with mitigation strategies that shape the robust use of these mobile quantum sensors for biosensing and nanoscale science.

\end{abstract}

\maketitle

\vspace{-1em}
\section*{Introduction}
\vspace{-1em}

Electronic spins hosted by solid-state point defects provide an atom-like platform for quantum control, with the nitrogen--vacancy (NV) center in diamond, most commonly operated in its negatively charged state (NV$^{-}$), serving as a widely studied and well-established system~\cite{wolfowicz2021quantum}. Owing to their long spin coherence times (up to $\sim$ms) under ambient conditions~\cite{balasubramanian2009ultralong}, NV centers have become central to quantum technologies, particularly for nanoscale sensing~\cite{childress2014atom}. Their stable fluorescence, tightly linked to the spin state, enables optical initialization and readout of NV spins, allowing electron spin resonance to be detected optically under microwave driving~\cite{bucher2019quantum,levine2019principles}. This approach, known as optically detected magnetic resonance (ODMR), has established NV centers as nanoscale magnetic sensors~\cite{taylor2008high,grinolds2013nanoscale}, with sensitivity extending to the single-molecule level~\cite{du2024single}. NV-based sensing has since expanded beyond magnetometry to include temperature~\cite{kucsko2013nanometre}, electric fields and strain~\cite{dolde2011electric}, and local chemical variables~\cite{fujisaku2019ph}, providing powerful tools for condensed-matter physics and materials science~\cite{casola2018probing,basu2025diamond} and, increasingly, for chemical and biological systems~\cite{mzyk2021diamond,aslam2023quantum}.

NV centers in bulk diamond have enabled much of the foundational progress in NV-based quantum sensing, while hosting these defects in diamond nanoparticles, hereafter referred to as nanodiamonds (NDs), realizes mobile quantum nanosensors. As freely deliverable probes, NDs can access intracellular and otherwise inaccessible nanoscale environments where macroscopic crystals are impractical~\cite{feng2022recent,yingke2022advscireview}. Their non-blinking, photostable fluorescence and chemical robustness, combined with low cytotoxicity, make NDs well suited for sustained operation in biological systems~\cite{vaijayanthimala2012long,mochalin2020properties}. Embedded NV centers further enable multimodal sensing in living systems, including intracellular nanothermometry~\cite{fujiwara2021diamond}, detection of specific molecular and redox-active species~\cite{rendler2017optical,mzyk2022relaxometry}, and nanoscale dynamics~\cite{Cui2022NanoLett}. Reducing diamond to the nanoscale, however, fundamentally reshapes the material landscape, imposing constraints on NV spin properties, charge-state stability, and sensing fidelity that are central to this Perspective.

\begin{figure*}[th!]
 \centering
 \includegraphics[width=\textwidth]{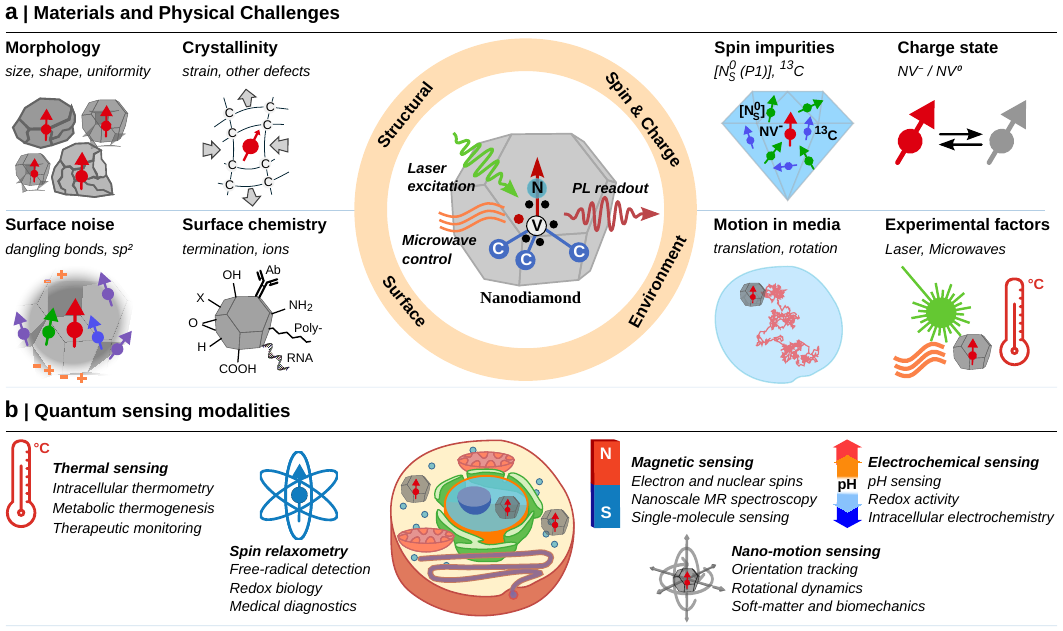}
 \vspace{-2em}
 \caption{\textbf{Bottlenecks and sensing opportunities of NDs hosting nitrogen--vacancy (NV) centers.}
\textbf{a.} Key factors limiting NV-based quantum sensing performance using NDs, grouped into four broad classes: structural disorder, surface-related effects, spin and charge-state control, and environmental influences. The central schematic illustrates a ND hosting an NV center and the core sensing principle based on optical initialization and photoluminescence (PL) readout of the NV spin, with microwave control applied where required.
\textbf{b.} Representative sensing opportunities enabled by NDs, highlighting their role as mobile probes in complex biological and nanoscale   environments.}
 \label{figure1}
\end{figure*}

At the nanoscale, mechanisms limiting NV performance that are well characterized in bulk diamond become intertwined and far less systematic. In bulk diamond, NV spin dynamics are primarily governed by volumetric spin impurities, with surface effects entering primarily from a single interface~\cite{bauch2020decoherence,barry2020sensitivity}. By contrast, in NDs, reduced dimensions place NV centers in close proximity to a three-dimensional (3D) surface environment, introducing multiple competing noise channels. In biological sensing contexts, background fluorescence necessitates high photon flux and often favors ensemble NV operation, while stabilizing the NV$^{-}$ state typically requires elevated concentrations of nitrogen impurities acting as electron donors, which are paramagnetic in nature and introduce unavoidable magnetic noise. As particle size is reduced, surface-related magnetic dipoles become increasingly important as NV--surface distances decrease~\cite{tetienne2013spin}. 
Nanoscale confinement further enhances lattice distortions (strain fields) and surface charge–induced electric-field fluctuations, both of which degrade NV spin properties~\cite{bauch2018ultralong,blinder2024reducing,chrostoski2022surface}. 
Surface charges also govern the charge-state stability of near-surface NV centers~\cite{hauf2011chemical,yamano2017charge,rondin2010surface}. Altogether, these effects limit sensing fidelity in NDs and produce a substantial sensitivity gap relative to bulk diamond~\cite{fujiwara2021diamond}. Particle-to-particle variability, arising from top-down synthesis routes with limited morphological control~\cite{https://doi.org/10.1002/ppsc.201900009}, further complicates quantitative sensing and reproducibility across experiments~\cite{mzyk2022relaxometry}. These concurrent bottlenecks, together with the sensing opportunities enabled by NDs, are summarized in Fig.~\ref{figure1}.

A growing body of work demonstrates that several limitations of NDs are not immutable. Advances in synthesis routes~\cite{alkahtani2019growth,alkahtanihydrothermal}, NV creation protocols~\cite{bao2025quantumgrade}, impurity control~\cite{oshimi2023quantum,PhysRevApplied.20.044045}, and surface treatments~\cite{zvi2025engineering,barzegaramiriolya2025functionalized,alkahtani2025advanced} have produced NDs with markedly improved spin and optical properties, in some cases approaching bulk-like regimes, albeit often with accompanying trade-offs.
In the recent literature, such materials are frequently described as ``quantum-grade'' NDs~\cite{alkahtanihydrothermal,bao2025quantumgrade,oshimi2023quantum,alkahtani2025advanced,copak2025linker}. In this Perspective, the term ``quantum-grade'' denotes ND platforms in which application-optimized NV behavior is sufficiently stable, reproducible, and physically interpretable, enabling quantitative NV-based quantum sensing within a materials-limited noise framework rather than by comparison to bulk diamond performance. Motivated by these requirements, this Perspective consolidates the physical mechanisms that govern how nanoscale structure and environment constrain NV behavior in NDs. The article is structured as follows. We first outline the principles of NV-based sensing in nanoscale environments, followed by a materials-focused discussion of ND production, including NV formation, surface chemical control, and morphology as they relate to NV behavior. Building on this foundation, we then analyze NV spin dynamics in the context of dominant magnetic, electric-field, and strain-related noise mechanisms, and conclude by identifying key considerations for achieving reliable and reproducible ND-based quantum sensing under realistic, practical scenarios.

\begin{figure*}[th!]
 \centering
 \includegraphics[width=\textwidth]{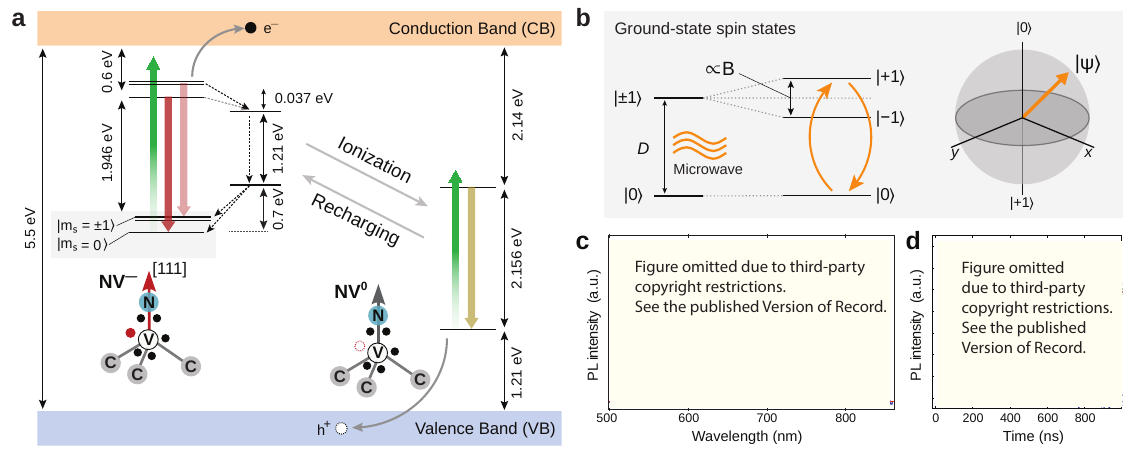}
 \vspace{-2em}
 \caption{\textbf{Fundamentals of nitrogen--vacancy (NV) centers in diamond.}
\textbf{a.} Electronic structure of NV defect states within the diamond band gap, illustrating optical excitation, radiative decay, and spin-dependent intersystem crossing in the NV$^{-}$ state.
\textbf{b.} Ground-state spin structure of NV$^{-}$. The spin-triplet ($S=1$) ground state enables coherent microwave control via the $\ket{0}\leftrightarrow\ket{\pm1}$ transitions, illustrated by the Bloch-sphere representation.
\textbf{c.} Normalized PL spectra of NV$^{0}$ and NV$^{-}$. Adapted from Ref.~\citenum{rondin2010surface} with permission. Copyrighted by the American Physical Society.
\textbf{d.} Spin-dependent PL dynamics under optical excitation, forming the basis of optical initialization and readout of the NV spin. Adapted from Ref.~\citenum{janik2013quantumzeno} with permission. Copyrighted by the American Physical Society.
} 
 \label{figure2-a}
\end{figure*}

\vspace{-1em}
\section*{Principles of NV-based Quantum Sensing}
\vspace{-1em}

Quantum sensing with NV centers in diamond has been extensively reviewed, encompassing their electronic structure, spin dynamics, and sensing modalities~\cite{doherty2013nitrogen,schirhagl2014nitrogen,abe2018tutorial,levine2019principles,bucher2019quantum}. Here, we highlight only the essential principles needed to understand how ND material characteristics govern NV spin properties and sensing performance, as examined in subsequent sections.

\vspace{-1em}
\subsection*{Fundamentals of NV Centers in Diamond}
\vspace{-1em}

In the diamond lattice, a substitutional nitrogen atom adjacent to a vacancy forms an NV center with $C_{3v}$ symmetry. The NV can exhibit multiple charge states. Among these, the negatively charged NV$^{-}$ state is central to quantum sensing, as it uniquely combines optical addressability with a spin-triplet electronic ground state ($S=1$) (Fig.~\ref{figure2-a}a,b). Under optical excitation, spin-selective relaxation through a metastable singlet manifold produces a contrast in photoluminescence (PL) between the $\ket{m_s=0}$ and $\ket{m_s=\pm1}$ spin sublevels (Fig.~\ref{figure2-a}c,d), enabling optical readout of the NV spin state. The same process preferentially polarizes the system into the $\ket{m_s=0}$ sublevel and, together with long longitudinal relaxation times ($T_1$ of a few ms in bulk diamond~\cite{jarmola2012temperature}), yields high spin polarization at room temperature~\cite{patel2024single}. In addition, the ground-state spin sublevels, separated by the zero-field splitting ($D \approx 2.87$~GHz), can be coherently driven by microwave fields (Fig.~\ref{figure2-a}b), allowing quantum superposition states to be prepared and manipulated at room temperature, with coherence times ($T_2$) up to $\sim$ms in bulk diamond~\cite{balasubramanian2009ultralong}. Together, efficient optical initialization, coherent spin control, and optical readout form the basis of ODMR, enabling sensitive detection of magnetic fields down to the single-spin level~\cite{grinolds2013nanoscale}, sub-kelvin temperature shifts~\cite{kucsko2013nanometre}, and local strain or electric-field variations~\cite{dolde2011electric}, with nanoscale spatial resolution under ambient conditions.

NV-based quantum sensing exploits the NV$^{-}$ state; however, it can coexist with the optically active neutral NV (NV$^{0}$) state (Fig.~\ref{figure2-a}a,c), as well as a positively charged NV$^{+}$ state, which is not optically active. Optical excitation drives metastable interconversion between NV$^{-}$ and NV$^{0}$ via electron and hole exchange with the diamond bands, with ionization and recharging rates depending on both excitation wavelength and power~\cite{NJP-aslam-2013}. This leads to time-dependent fluctuations in the charge state of individual NV centers, as observed in single-NV experiments, which can affect spin-initialization fidelity~\cite{PhysRevLett.122.076101} and spin-relaxation dynamics~\cite{yamano2017charge}. In ensemble measurements, however, these transient dynamics are largely averaged out, and the system is instead characterized by a steady-state NV$^{-}$ population determined by excitation conditions.

\begin{figure*}[th!]
 \centering
 \includegraphics[width=\textwidth]{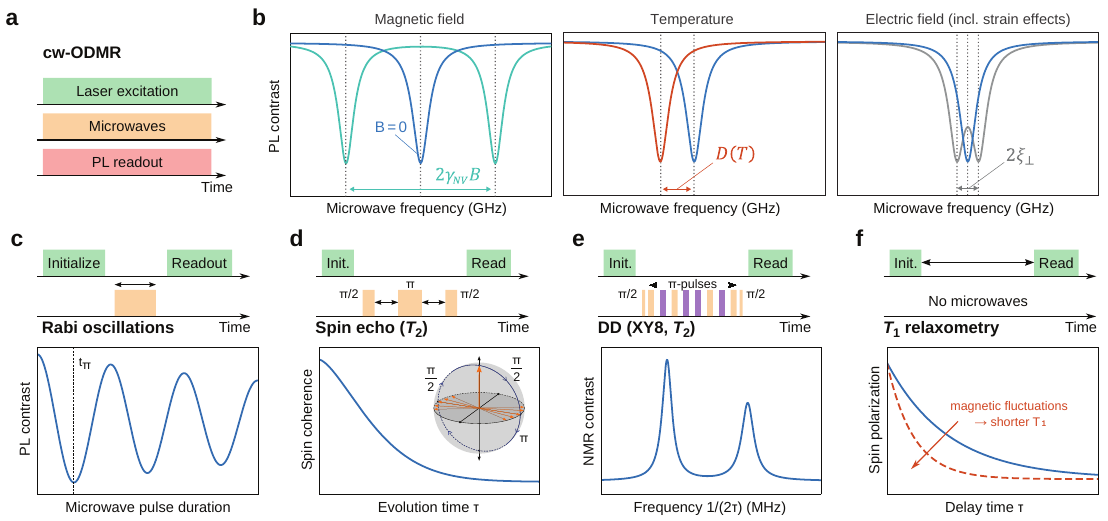}
 % \vspace{-2em}
 \caption{\textbf{Sensing principles of NV centers in diamond.}
\textbf{a.} Continuous-wave ODMR (cw-ODMR), in which laser excitation, microwave driving, and PL readout are applied simultaneously.
\textbf{b.} Representative cw-ODMR spectra illustrating sensitivity to magnetic field, temperature, and electric field, including strain contributions.
\textbf{c--f.} Representative pulsed sensing protocols, including Rabi oscillations (\textbf{c}), spin-echo measurements of the coherence time $T_2$ (\textbf{d}), and dynamical-decoupling sequences (e.g., XY8) enabling frequency-selective detection of nuclear magnetic resonance (NMR) signals (\textbf{e}).
\textbf{f.} Spin--lattice relaxation ($T_1$) measurements, sensitive to high-frequency magnetic fluctuations. See Table~\ref{tab:NVprotocols} for a summary of sensing protocols and their spectral selectivity.
}
 \label{figure2-b}
\end{figure*}

% \clearpage
The interaction of the NV spin with its local environment, encompassing material-intrinsic and external perturbations
relevant to sensing applications, is captured by the groundstate Hamiltonian (Eq.~\ref{eq:hamitonian})~\cite{PhysRevB.93.024305,dolde2011electric}: 

\vspace{-3em}
\begin{widetext}
\begin{align}
\frac{\mathcal{H}}{h} =&\;
\underbrace{D(T) S_z^2}_{\mathcal{H}_\text{ZF}~(\text{temperature})}
+ \underbrace{\gamma_{NV}\, \mathbf{B} \cdot \mathbf{S}}_{\mathcal{H}_\text{Zeeman}~(\text{magnetic})}
+ \underbrace{\mathbf{S} \cdot \mathbf{A} \cdot \mathbf{I}
+ P I_z^2
+ \gamma_N\, \mathbf{B} \cdot \mathbf{I}}_{\mathcal{H}_\text{nuclear}} \notag\\
& + \underbrace{
d_{\parallel} \Pi_z S_z^2 
- d_{\perp} \!\left[
\Pi_x (S_x S_y + S_y S_x)
+ \Pi_y (S_x^2 - S_y^2)
\right]
}_{\mathcal{H}_\text{ZF},~\mathcal{H}_{\mathcal{E},S}~(\text{electric / strain coupling})}
\notag\\[2mm]
\text{with}\quad
&\Pi_i = \mathcal{E}_i + M_i, \qquad i \in \{x,y,z\},
\qquad
\Pi_\perp = \sqrt{\Pi_x^2 + \Pi_y^2},
\notag\\
&\xi_\parallel = d_\parallel \Pi_z,
\qquad
\xi_\perp = d_\perp \Pi_\perp.
\label{eq:hamitonian}
\end{align} 
\end{widetext}

The dominant zero-field splitting term $D(T)$ arises from intrinsic spin--spin interactions and sets the NV sensitivity to temperature. External magnetic fields couple through the Zeeman interaction with $\gamma_{\mathrm{NV}} \simeq 28~\mathrm{GHz/T}$, while hyperfine and quadrupolar interactions with nearby nuclear spins, primarily $^{14}$N and $^{13}$C, introduce additional spectral structure. Non-magnetic perturbations enter through the effective fields $\Pi_i$, which combine electric fields $\mathcal{E}_i$ and strain-induced contributions $M_i$, as the latter couple to the NV spin through the same linear Stark interaction as $\mathcal{E}_i$~\cite{PhysRevB.93.024305,dolde2011electric}. 
The corresponding axial and transverse couplings, $\xi_\parallel = d_\parallel \Pi_z$ and $\xi_\perp = d_\perp \Pi_\perp$, with $d_{\perp}=17~\mathrm{Hz\,cm/V}$ and $d_{\parallel}=0.35~\mathrm{Hz\,cm/V}$~\cite{van1990electric}, cause shifts and splittings of the NV spin sublevels, analogous to temperature- and magnetic-field-induced effects, and play an important role in determining spin properties and spectral linewidths. In bulk diamond, strain and electric-field disorder are typically small and well controlled, enabling narrow resonance linewidths and long coherence times~\cite{balasubramanian2009ultralong}. In NDs, by contrast, lattice strain arising from nanoscale crystallinity and morphological irregularities, together with electric-field disorder from surface charges, often dominate NV behavior. These perturbations lead to inhomogeneous zero-field splitting, broadened spin resonances, reduced spin relaxation times, and pronounced particle-to-particle variability, as discussed in the following sections.

\begin{table*}[t]
\caption{\textbf{Summary of NV sensing protocols and spectral selectivity.} Different NV measurement schemes probe distinct frequency ranges and environmental couplings through their characteristic spin dynamics. Representative protocols and signals are shown in Fig.~\ref{figure2-b}.}
\label{tab:NVprotocols}
\begin{ruledtabular}
\begin{tabular}{
p{2.2cm}
p{2.2cm}
p{1.8cm}
p{4.0cm}
p{1.3cm}
p{4.2cm}
}
\textbf{Protocol} &
\textbf{Sequence} &
\textbf{Frequency range} &
\textbf{Dominant noise sensitivity} &
\textbf{Spin time} &
\textbf{Sensing modalities} \\

cw-ODMR &
Continuous laser + MW sweep &
DC--kHz &
Optical and MW power broadening &
-- &
DC magnetometry, temperature, strain, electric fields, rotation \\

Rabi &
Laser--MW--laser &
On-resonance &
MW amplitude and detuning &
$T_2^*$ &
Spin control calibration ($\pi$, $\pi/2$ pulses), MW-field mapping \\

Ramsey &
$\pi/2$--$\tau$--$\pi/2$ &
DC--few kHz &
Quasi-static magnetic, electric, and strain noise &
$T_2^*$ &
DC magnetometry, temperature, strain, electric fields \\

Spin echo (Hahn) &
$\pi/2$--$\tau$--$\pi$--$\tau$--$\pi/2$ &
kHz--MHz &
Time-varying magnetic and electric noise &
$T_2$ &
AC magnetometry, noise spectroscopy \\

Dynamical decoupling (CPMG, XY8) &
$\pi/2$--($\pi$)$^n$--$\pi/2$ &
Tunable (kHz--MHz) &
Spectrally selective magnetic and electric noise &
$T_2$ & %Extended
High-resolution NV--NMR, noise spectroscopy \\

$T_1$ relaxation  & %relaxometry
Laser--dark--laser &
GHz &
High-frequency surface and environmental noise &
$T_1$ &
Paramagnetic species, broadband noise spectroscopy, pH \\
\end{tabular}
\end{ruledtabular}
\end{table*}

% \vspace{-1em}
\subsection*{Quantum Sensing Modalities and Protocols}
\vspace{-1em}

The NV Hamiltonian can be probed using a range of experimental protocols that access distinct frequency regimes and environmental couplings (Table~\ref{tab:NVprotocols} and Fig.~\ref{figure2-b}). These primarily take the form of continuous-wave (cw-) ODMR (Fig.~\ref{figure2-b}a,b) or time-resolved pump--probe measurements, including pulsed ODMR (Fig.~\ref{figure2-b}c--e) and all-optical $T_1$ relaxometry (Fig.~\ref{figure2-b}f), as outlined below.

In cw-ODMR (Fig.~\ref{figure2-b}a), optical pumping, microwave excitation, and PL readout are applied simultaneously. A frequency sweep of the microwave field produces resonances around $D$, and an external magnetic field aligned along the NV axis induces Zeeman splitting (Fig.~\ref{figure2-a}b). Temperature effects appear as shifts of $D$~\cite{fujiwara2021diamond}, while weak splittings of $2\xi_\perp$ reflect lattice strain and/or static electric fields arising from the intrinsically weak Stark effect~\cite{dolde2011electric}. In this measurement mode, the smallest detectable frequency shift is determined by the detected photon rate $I_0$, contrast $C$, and the resonance linewidth $\Delta\nu$, with the sensitivity scaling approximately as $\eta_{\mathrm{cw}} \propto \Delta\nu /(C\sqrt{I_0})$~\cite{dreau2011avoiding}. 
Although the fundamental lower bound of $\Delta\nu$ is set by the inhomogeneous dephasing time $T_2^*$, such that $\Delta\nu \sim 1/(\pi T_2^*)$, continuous optical repolarization and microwave driving introduce power broadening of $\Delta\nu$, while also reducing $C$~\cite{dreau2011avoiding}. 
In NDs, enhanced noise sources further broaden the resonances, strongly degrading sensitivity relative to bulk diamond~\cite{fujiwara2021diamond}, and random particle orientations preclude reliable vector magnetometry, particularly in biological environments.

Advanced time-resolved protocols temporally separate spin manipulation from optical pumping and readout (Fig.~\ref{figure2-b}c--f), enabling access to intrinsic NV spin dynamics. Rabi oscillations (Fig.~\ref{figure2-b}c) establish coherent microwave control of the NV spin between the $\ket{0}$ and $\ket{\pm1}$ states, typically under an applied bias magnetic field (Fig.~\ref{figure2-a}b). Building on this control, Ramsey sequences employ a pair of $\pi/2$ pulses to create and read out a spin superposition that accumulates phase during a free-evolution interval $\tau$ from quasi-static perturbations (DC--few kHz), directly probing $T_2^*$ and providing a narrower effective bandwidth and higher sensitivity than cw-ODMR. Introducing a refocusing $\pi$ pulse yields the spin-echo sequence (Fig.~\ref{figure2-b}d), which suppresses slowly varying noise, extends coherence to $T_2$, and enables sensitivity to time-dependent fields. Multiple refocusing pulses, as implemented in dynamical-decoupling sequences such as XY8 (Fig.~\ref{figure2-b}e), further prolong coherence and enhance frequency selectivity, forming the basis of NV-detected NMR~\cite{du2024single}. In NDs, short $T_2$ times and magnetic-field alignment constraints limit the applicability of pulsed protocols, motivating alternative strategies to improve coherence-based sensing~\cite{holzgrafe2020nanoscale}.
Spin--lattice relaxation ($T_1$ relaxometry) provides an all-optical sensing modality that probes population transfer between the $\ket{0}$ and $\ket{\pm1}$ states and is therefore sensitive to environmental fluctuations near the NV transition frequency (Fig.~\ref{figure2-b}f). Because it does not require microwave control or magnetic-field alignment, $T_1$ relaxometry is particularly attractive for ND-based sensing and is widely used to detect paramagnetic species in practical applications~\cite{mzyk2022relaxometry}. 
Yet, reduced $T_1$ and pronounced particle-to-particle variability in NDs complicate quantitative interpretation, with additional environmental contributions arising from changes in surface charge (e.g., under pH variations) and ionic conditions~\cite{freire2023role,fujisaku2019ph}.

These protocols link the NV spin Hamiltonian to experimentally accessible observables. In NDs, deviations from bulk-diamond benchmarks in spectral linewidths and characteristic spin times directly reflect enhanced noise sources at the nanoscale. The following sections examine the materials origins of these effects and discuss strategies to systematically improve ND-based quantum sensing performance.

\begin{figure*}[t!]
 \centering
 \includegraphics[width=\textwidth]{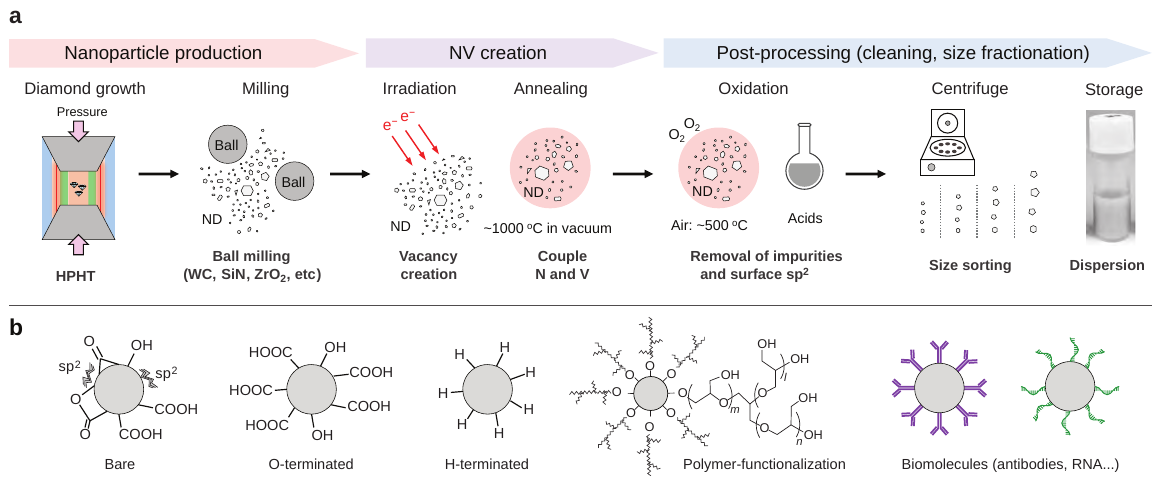}
 \vspace{-2em}
\caption{\textbf{Overview of ND fabrication and chemical control for NV-based quantum sensing.}
\textbf{a.} Commonly used workflow for ND fabrication, including nanoparticle synthesis, NV creation, and post-processing steps. 
\textbf{b.} Representative ND surface states, including bare surfaces with residual $sp^2$-like carbon, oxygen (O)- and hydrogen (H)-terminated surfaces, and polymer- or biomolecule-functionalized NDs for colloidal stabilization and biofunctionalization.
 }
 \label{fig3}
\end{figure*}

\section*{Nanodiamond Fabrication and Spin-Hosting Material Properties}
\vspace{-1em}

\begin{figure}[th!]
 \centering
 \includegraphics[width=\columnwidth]{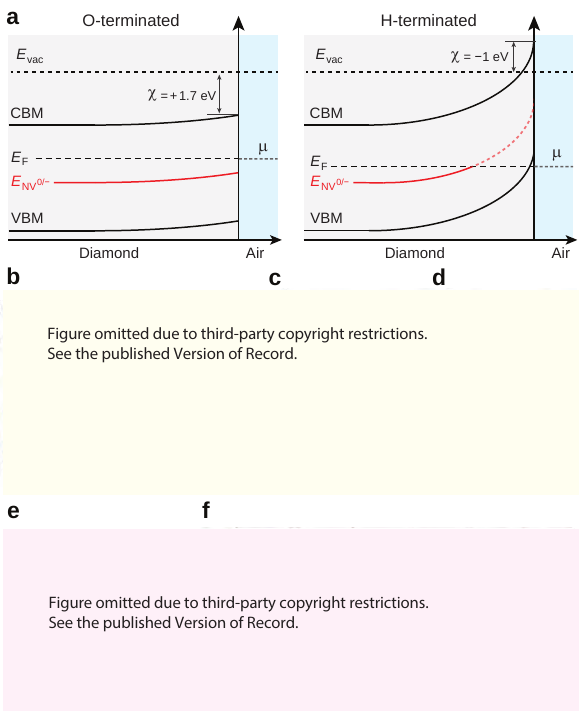}
 \vspace{-2em}
 \caption{\small
 \textbf{NV charge-state control through surface electrostatics and electrochemical modulation.} 
   \textbf{a.} Schematic band diagrams of O- and H-terminated diamond surfaces in contact with air or aqueous environments, showing the vacuum level ($E_{\mathrm{vac}}$), electron affinity $\chi$, band bending, the conduction band minimum (CBM) and valence band maximum (VBM), Fermi level ($E_F$) alignment with the environmental chemical potential ($\mu$), and the NV$^{0/-}$ charge-transition level ($E_{\mathrm{NV}^{0/-}}$).
    \textbf{b–d.} Simulated band diagrams for H-terminated diamond at two nitrogen implantation doses of $10^{12}$ and $10^{13}$ cm$^{-2}$, showing the implantation profile and depth-dependent alignment of the VBM (solid) and NV charge-transition levels (NV$^{+/0}$, dashed; NV$^{0/-}$, dotted), together with PL intensity difference spectra under electrochemical gating (${+}$0.5 V and ${-}$0.5 V). Adapted with permission from Ref. \citenum{grotz2012charge}.
    \textbf{e-f.} Simulated 2D map of the VBM near the ITO/ND interface, highlighting regions where $E_{\mathrm{NV}^{0/-}} < E_F$ (green) for NV$^{-}$ stabilization; corresponding PL measurements show voltage-induced modulation of the NV charge state in ND ensembles with single NV centers. Adapted with permission from Ref.~\citenum{karaveli2016modulation}.
}
\label{figure-charge-state}
\end{figure}

NV-based sensing performance in NDs is strongly influenced by material-dependent noise sources, including lattice strain, proximity to charges, and spin impurities, all of which are amplified by nanoscale confinement.
ND fabrication typically involves nanoparticle synthesis and NV formation, followed by cleaning, size fractionation, and, in many cases, deliberate surface modification to improve reproducibility and sensing performance (Fig.~\ref{fig3}a,b). 
These fabrication steps define a set of spin-hosting material properties, including the surface electronic structure that governs band bending of defect levels and, in turn, NV charge-state stability (Fig.~\ref{figure-charge-state}a), as well as particle morphology. Together, these factors establish the local environment experienced by NV spins. Here, we review ND production, surface chemistry, and particle morphology from an NV-based quantum-sensing perspective, providing the materials context for the noise mechanisms and mitigation strategies discussed in subsequent sections.

\vspace{-1em}
\subsection*{Nanodiamond Synthesis, NV Creation, and Post-processing}
\vspace{-1em}

Diamond nanoparticles are most commonly obtained via top-down milling of bulk diamond, detonation-based synthesis, or emerging bottom-up approaches. Among these, top-down milling routes are most widely used in quantum-sensing studies. In this approach, bulk diamond grown by high-pressure high-temperature (HPHT) or chemical vapor deposition (CVD) methods~\cite{Nebel31122023,teraji2025reviewRSC} is fragmented through high-energy collisions with hard milling media such as tungsten carbide~\cite{boudou2009high}, silicon nitride~\cite{doi:10.1021/acsomega.8b02067,PhysRevApplied.20.044045}, or zirconia~\cite{OSAWA20072018}. Milling unavoidably introduces surface contamination and non-diamond carbon, necessitating further treatments that also shape NV properties and colloidal stability, as discussed below. Detonation synthesis provides access to ultrasmall NDs ($<10$~nm) but typically yields inferior NV spin properties due to heterogeneous surfaces and high strain. Additionally, the probability of incorporating NV centers remains low, often on the order of one per 10$^3$-10$^4$ particles~\cite{smith2010effects}. Recent work, however, has demonstrated improved NV performance in detonation NDs~\cite{likhachev2023odmr,sotoma2018enrichment,doi:10.1021/acsnano.8b09383,10.1063/5.0201154}.

Although NV centers can form incidentally during ND synthesis~\cite{schroder2011ultrabright,Schroder:12,fujiwara2019monitoring}, deliberate defect engineering is generally required to achieve bright PL and robust sensing performance. This process involves incorporation of substitutional nitrogen (N$_\mathrm{S}$) and generation of lattice vacancies, followed by thermal diffusion to form NV centers~\cite{Smith2019nanophotonics}. Stabilization of the NV$^{-}$ charge state is governed by donor--acceptor charge transfer, most commonly involving neutral N$_\mathrm{S}$ (N$_\mathrm{S}^{0}$; commonly referred to as P1 centers)~\cite{rondin2010surface}:

\vspace{-1em}
\begin{equation}
\mathrm{NV}^{0} + \mathrm{N}{_\mathrm{S}}^{0} \rightleftharpoons \mathrm{NV}^{-} + \mathrm{N}{_\mathrm{S}}^{+}.
\label{eq:N-NV-charge exchange}
\end{equation}

Balancing the N$_\mathrm{S}$ concentration, [N$_\mathrm{S}$], and vacancy density therefore maximizes the NV$^{-}$ yield while limiting residual P1 centers (paramagnetic impurities) and NV$^{0}$ populations, both of which degrade sensing performance. 
In practice, [N$_\mathrm{S}$] is typically set during bulk diamond growth prior to milling, either through nitrogen getters in HPHT synthesis or intentional incorporation during CVD growth~\cite{https://doi.org/10.1002/admi.201901408,teraji2025reviewRSC}, with post-growth nitrogen implantation also possible~\cite{Smith2019nanophotonics}. Commercial type-Ib NDs milled from HPHT-grown diamonds typically contain [N$_\mathrm{S}$] of $\sim$50--500~ppm~\cite{teraji2025reviewRSC}. Vacancies are introduced by electron irradiation, ion implantation, or laser writing, with irradiation most commonly employed owing to its efficiency and scalability. Subsequent annealing at 600--1000~\si{\degreeCelsius} enables vacancy diffusion and NV formation, yielding typical [NV] of $\sim$3~ppm and nitrogen-to-NV$^{-}$ conversion efficiencies of 1--20\%~\cite{Luo_2022,shenderova2019synthesis,PhysRevApplied.10.034044,Edmonds_2021,qubs6010002}. Detailed comparisons of NV creation protocols are reviewed elsewhere~\cite{Smith2019nanophotonics,shenderova2019synthesis,teraji2025reviewRSC}.

Achieving high-quality NV ensembles requires careful optimization. Insufficient vacancy creation limits NV yield at high [N$_\mathrm{S}$], whereas excessive irradiation depletes donors and introduces compensating defects, such as acceptor states, that favor NV$^{0}$ formation~\cite{collins2002fermi,waldermann2007creating,acosta2009diamonds,teraji2025reviewRSC,Luo_2022}. Quantitative studies in bulk diamond show near-complete NV$^{-}$ stabilization for [NV]/[N$_\mathrm{S}^{0}$] ratios up to $\sim$20\%, beyond which the NV$^{0}$ fraction gradually increases~\cite{APL2021NVstate}. 
Similarly, in NDs, increasing irradiation fluence enhances the NV$^{-}$ fraction; however, it saturates well below unity (typically $\sim$70--75\%) and decreases rapidly at higher damage levels~\cite{apra2024creation}. 
This behavior reflects the enhanced susceptibility of NDs to lattice damage, which promotes surface-induced electronic depletion and the formation of compensating acceptor states~\cite{santori2009vertical}, suppressing NV$^{-}$ stability well before bulk-like limits are reached. These effects reduce the NV$^{-}$ fraction with decreasing particle size~\cite{rondin2010surface} and lead to pronounced particle-to-particle variability, often correlated with shape irregularities.

ND synthesis and NV creation both perturb the particle surface, introducing contamination and non-diamond carbon, particularly graphitic species. 
These processes also promote aggregation, necessitating repeated cleaning and deagglomeration steps~\cite{krueger2008structure}. Oxidative acid treatments, typically using triacid or \ce{HNO_3}/\ce{H_2SO_4} mixtures at $\sim$100~\si{\degreeCelsius}, have therefore become standard post-processing protocols~\cite{Arnault2019,boudou2009high,doi:10.1021/acsomega.8b02067,PhysRevApplied.20.044045,OSAWA20072018}. Beyond cleaning, these treatments define the surface chemistry by introducing oxygen-containing groups (e.g., --OH, --COOH), which govern NV performance and colloidal stability while providing scaffolds for subsequent functionalization~\cite{sreeramoju2021surface,terada2022simple,sivtseva2025effect,LI2021725,jung2021surface}. Aerobic annealing at elevated temperatures ($\sim$500~\si{\degreeCelsius} in air) further improves surface quality, yielding terminations closer to bulk diamond and enhancing NV spin properties~\cite{doi:10.1021/jp506992c,tsukahara2019removing,shestakov2019advanced}. Accordingly, combined acid cleaning and aerobic oxidation are widely regarded as effective routes to chemically robust ND surfaces for quantum sensing~\cite{LI2021725}. More recently, a multistep post-treatment combining molten-salt oxidation with sequential acid and alkaline cleaning has been shown to promote morphological reshaping into faceted NDs with narrower size dispersions, while enhancing colloidal stability, PL brightness, and $T_1$~\cite{alkahtani2025advanced}.

Size fractionation represents a final critical step, as size dispersion directly translates into variability in NV properties. Centrifugation-based methods are commonly employed, typically yielding fractions with mean sizes from $\sim$20 to 500~nm~\cite{LI2018645}, although nominal size labels often mask broad distributions, motivating refined approaches such as viscosity-gradient centrifugation to achieve narrower and more reproducible size dispersions~\cite{finas2024efficiently}.

\subsection*{Surface treatment and chemical control}
\vspace{-1em}

At the nanoscale, surface states dominate the local noise environment and enable molecular adsorption, thereby influencing NV properties, colloidal behavior, and overall sensing performance. Surface control is therefore essential for reliable sensing (Fig.~\ref{fig3}b). A primary strategy is surface termination, which passivates the ND surface to ensure chemical robustness and colloidal stability in liquid-phase and biological environments. At the same time, surface termination defines the electrostatic boundary conditions that induce band bending of NV defect levels relative to the diamond Fermi level ($E_\mathrm{F}$), thereby governing NV charge-state stability (Fig.~\ref{figure-charge-state}a)~\cite{hauf2011chemical}.

Oxidative post-processing typically yields oxygen (O)-terminated ND surfaces bearing --OH and --COOH groups. O termination (electron affinity: $\chi \approx +1.7$~eV) results in slight upward band bending, i.e., near-flat band conditions, favoring the NV$^{-}$ charge state~\cite{hauf2011chemical,yamano2017charge}. This leads to enhanced PL~\cite{yamano2017charge,kim2014effect,laube2019controlling,apra2024creation}, aided by reduced surface electron trapping associated with improved surface crystallinity~\cite{rondin2010surface}. 
O-terminated NDs also exhibit strongly negative zeta potentials ($\zeta \lesssim -30$~mV) and stable colloidal dispersions over a broad pH range (approximately pH~4--10), making them well suited for liquid-phase sensing~\cite{fujisaku2019ph}. Hydrogen (H)-terminated surfaces, characteristic of as-grown CVD diamond, represent a contrasting surface state. C--H surface dipoles lower the electron affinity ($\chi \approx -1$~eV), promote electron transfer to adsorbates, and induce strong upward band bending that favors NV$^{0}$ formation near the surface~\cite{hauf2011chemical,petrakova2012luminescence}. H-terminated surfaces typically exhibit zeta potentials that are positive at low pH and reverse sign near physiological pH~\cite{doi:10.1126/science.1148841,STEHLIK2024100327}.
Between these extremes, mixed H/O surface terminations are predicted to optimize NV-based sensing~\cite{chou2017nitrogen} and have recently been shown to enhance NV spin properties in NDs via non-thermal plasma treatments~\cite{gulka2024surface}. Additional surface terminations, including nitrogen (N)-, amine (–NH$_2$)-, and fluorine (F)-based passivation, further extend the tunability of surface charge and reactivity~\cite{kawai2019jpcc,KONDO200848,shintani2017all}, underscoring surface chemistry as a key lever for mitigating surface-related noise and improving NV performance.

Beyond intrinsic surface termination, the NV charge state can be modulated by coupling between the diamond surface and the local environment. Electrochemical gating modifies the interfacial electrostatic boundary conditions, thereby shifting the diamond Fermi level relative to the NV charge transition levels and enabling dynamic control of the NV charge-state population~\cite{grotz2012charge} (Fig.~\ref{figure-charge-state}b-d). This effect underpins all-optical, charge-state–based imaging of voltage bias and electric fields, demonstrated both in electrochemical environments~\cite{mccloskey2022diamond} and under ambient conditions~\cite{NatCommun-12-2457}. These effects become increasingly relevant as the NV–surface distance decreases, a regime that is ubiquitous in NDs and particularly pronounced in single NV measurements~\cite{karaveli2016modulation} (Fig.~\ref{figure-charge-state}e-f).

While electrostatic stabilization through surface termination is often sufficient in simple aqueous environments, high ionic strength and nonspecific biomolecular adsorption in biofluids promote aggregation and protein corona formation, degrading colloidal stability~\cite{komatsu2023accounts}. Polymer coatings provide an effective route to steric stabilization under these conditions; for example, poly(glycerol)-grafted NDs exhibit excellent dispersion stability in complex biofluids while suppressing nonspecific protein adsorption~\cite{zou2020acsnano}. Tuning polymer charge and architecture further enables control over cellular uptake~\cite{https://doi.org/10.1002/adfm.202270125}. Beyond polymer passivation, conjugation with biomolecules enables site-specific targeting and sensing, such as NV-assisted lateral flow assays and measurements of receptor dynamics in living cells~\cite{miller2020spin,doi:10.1021/jacs.0c01191}. Recently, Copak \textit{et al.} demonstrated azide-terminated NDs enabling linker-free covalent DNA functionalization while preserving NV charge stability and $T_1$~\cite{copak2025linker}. As reliable localization and retention are prerequisites for meaningful quantum measurements, surface functionalization represents a key enabling step for ND-based biosensing.

\vspace{-1em}
\subsection*{ND Morphology}
\vspace{-1em}

NDs are typically produced by top-down milling processes, which offer limited control over particle morphology, including both size and shape. As a result, pronounced variability in NV properties persists even among samples with identical nominal sizes. Here, we outline how ND geometry fundamentally constrains NV-based sensing performance and contributes to particle-to-particle heterogeneity.

ND size is a fundamental parameter that sets the distance between NV centers, surface-related noise sources, and sensing targets, thereby defining both sensitivity and practical usability. For magnetic sensing modalities such as nanoscale NMR and $T_1$ relaxometry, signal strength follows the $r^{-3}$ decay of dipolar fields~\cite{taylor2008high,tetienne2013spin}. Reducing particle size therefore enhances signal strength through improved proximity but simultaneously increases coupling to surface-induced noise, which degrades spin coherence and ultimately limits sensitivity. Size dispersion further broadens NV--surface and NV--target distance distributions, increasing uncertainty in quantitative measurements~\cite{Sigaeva2022Small}. Although NV spin properties are not determined by size alone, improved size control is expected to enhance reproducibility across ND ensembles.

ND size also determines the number of NV centers per particle and thus the PL brightness~\cite{shenderova2019synthesis}, a critical constraint in biological environments with strong autofluorescence. For typical type-Ib NDs with [NV] $\sim$3~ppm, particles smaller than several tens of nanometers often yield insufficient signal for stable optical readout~\cite{fluorescent2018nanophotonics}. As a result, biological applications frequently favor larger NDs despite reduced spatial resolution and sensing proximity~\cite{oshimi2023quantum,kucsko2013nanometre,Cui2022NanoLett,fujiwara2020real}. Particle size further constrains biological accessibility; for example, transport through nuclear pores typically requires particles smaller than $\sim$30~nm~\cite{yang2014peptide}. Practical ND-based quantum sensing therefore requires careful size optimization to balance brightness, spin properties, and biological delivery constraints.

\begin{figure}[th!]
 \centering
 \includegraphics[width=8cm]{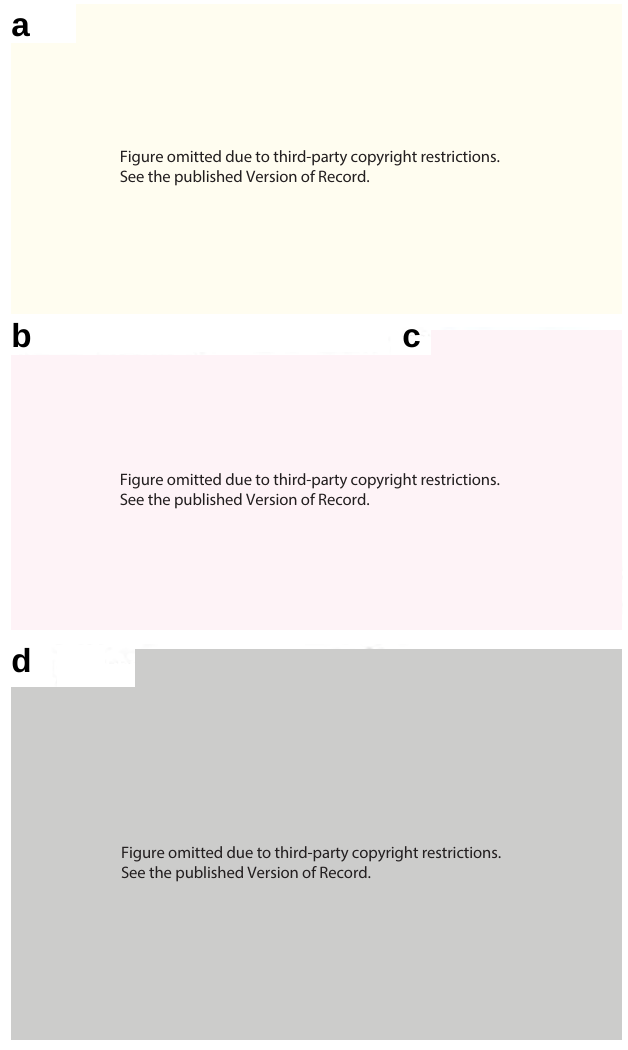}
 \vspace{-1em}
 \caption{\small \textbf{Shape anisotropy of NDs and its impact on PL heterogeneity.} 
\textbf{a.} ND size distributions obtained from AFM ($z$ and $x$--$y$) and DLS measurements, illustrating method-dependent size estimates for anisotropic particles. Inset: representative AFM 3D topography highlighting anisotropic particle geometry. Adapted with permission from Ref.~\citenum{ELDEMRDASH2023268}.
 \textbf{b.} Annular dark-field scanning TEM (ADF-STEM) and bright-field TEM (BF-TEM) images of NDs, validating BF-TEM contrast as a proxy for local thickness in quasi-3D morphology reconstruction.
 \textbf{c.} Representative reconstructed ND morphologies, revealing pronounced flake-like geometries.
 \textbf{d.} Machine-learning-based classification of ND shapes in Hu-moment space ($H_1$--$H_3$) and corresponding volume-averaged PL intensity, revealing three distinct shape populations with systematically different brightness. Panels b--d are adapted from Ref.~\citenum{haotin2023acsnano} with permission.
}
\label{figure5}
\end{figure}

In addition to size, top-down milling produces NDs with highly heterogeneous shapes that often deviate substantially from ideal spheres. Plate- or flake-like morphologies are common, leading to ambiguity in reported ``ND size'' depending on the measurement technique; for example, DLS predominantly reflects the lateral dimensions of plate-like particles, as confirmed by combined AFM and DLS measurements~\cite{ELDEMRDASH2023268} (Fig.~\ref{figure5}a). Recent work has directly linked ND shape to PL heterogeneity~\cite{https://doi.org/10.1002/ppsc.201900009,nano11102706,haotin2023acsnano}. Using TEM-based quasi-3D morphology reconstruction combined with unsupervised machine learning, Wen \textit{et al.} showed that a substantial fraction of NDs exhibit flake-like geometries and that PL brightness depends systematically on particle shape~\cite{nano11102706,haotin2023acsnano} (Fig.~\ref{figure5}b--d). This behavior was attributed to shape-dependent optical cavity and interference effects arising from the refractive-index contrast between diamond and the surrounding medium, identifying shape anisotropy as a major contributor to PL variability not captured by lower-dimensional analyses.

\begin{figure*}[th!]
 \centering
 \includegraphics[width=17cm]{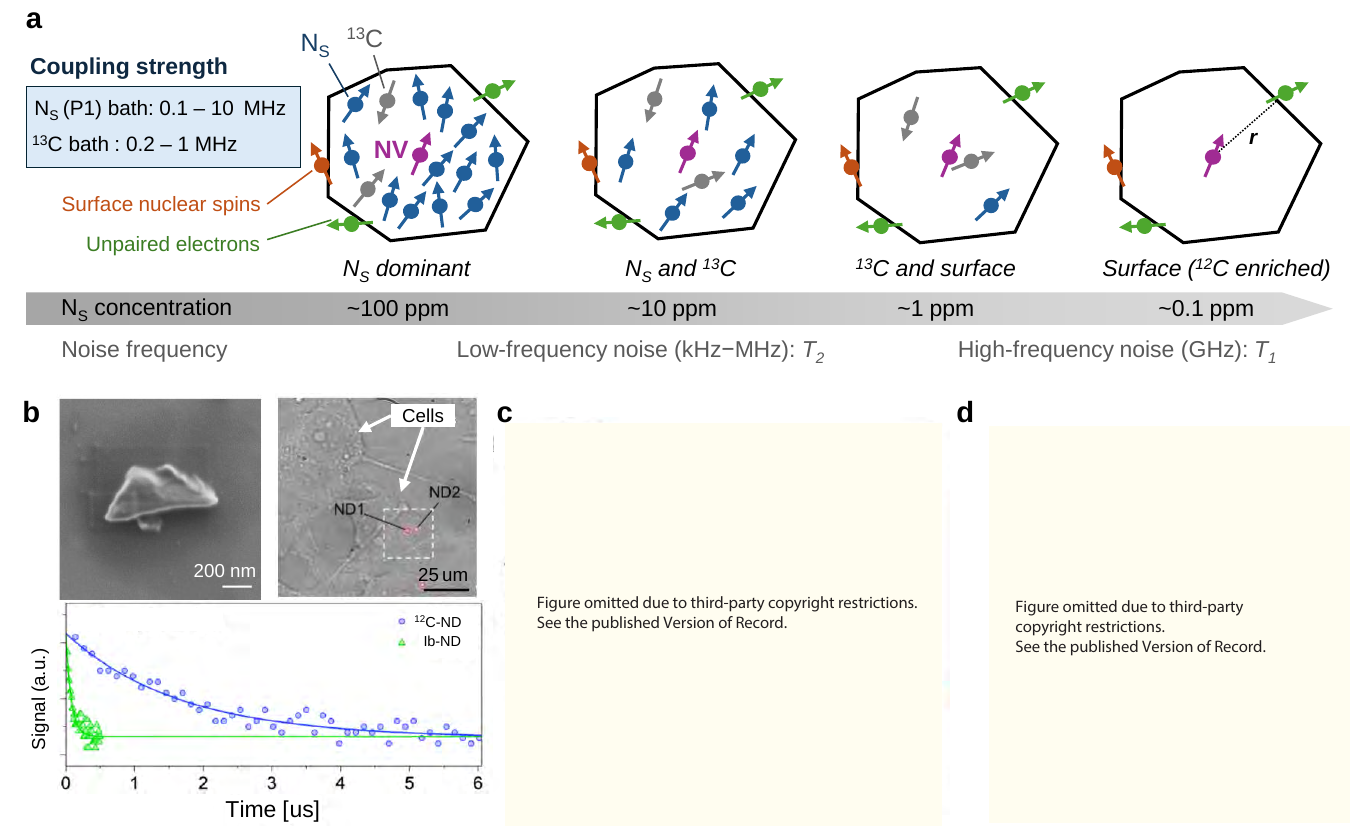}
 \vspace{-1em}
 \caption{\textbf{Noise sources and spin relaxation of NV centers in NDs.} 
\textbf{a.}  Schematic overview of dominant magnetic noise mechanisms as a function of substitutional nitrogen concentration ($[\mathrm{N_S}]$). The NV--surface distance $r$ highlights the strong distance dependence of surface-related noise. Typical magnetic interaction strengths and the characteristic frequency ranges governing $T_1$ and $T_2$ relaxation are indicated.
\textbf{b.} SEM image of a \ce{^{12}C}-enriched ND with high NV density, corresponding optical microscopy image of NDs internalized in cells, and representative $T_2$ spin-echo decay compared with type-Ib NDs. Adapted from Ref.~\citenum{oshimi2023quantum}, licensed under CC BY 4.0.
\textbf{c.} Particle size-dependent $T_1$ relaxation rates for type-Ib NDs, together with spherical-model simulations for a single NV located at the particle center or 3~nm below the surface, illustrating the dominance of surface-induced magnetic noise at small particle sizes. Adapted from Ref.~\citenum{tetienne2013spin} with permission.  Copyrighted by the American Physical Society.
\textbf{d.} TEM image of silica-coated NDs for mitigating surface-related noise. Reproduced with permission from Ref.~\citenum{barzegaramiriolya2025functionalized}. Copyright 2025 American Chemical Society. UoM-Si-3 denotes the sample name.
}
\label{Figure: spin impurities and surface noise}
\end{figure*}

Shape anisotropy is also expected to influence NV spin properties by setting the distribution of NV--surface distances within individual particles. Ong \textit{et al.} inferred such distributions from AFM-derived aspect ratios and suggested increased variability of $T_1$ and $T_2$ values~\cite{ong2017shape}. However, systematic experimental correlations between full 3D morphology and NV spin properties in NDs have yet to be established. Disentangling shape-induced effects from surface chemistry, spin impurities, and charge-state dynamics therefore remains an open challenge for robust ND-based quantum sensing.

\vspace{-1em}
\section*{Spin Relaxation and Noise Sources in NDs}
\vspace{-1em}

NV-based quantum sensing performance is governed by the characteristic NV spin relaxation times ($T_1$, $T_2$, and $T_2^*$; Table~\ref{tab:NVprotocols}), which encode the spectral density of the local noise environment. Noise at different frequencies couples selectively to the NV spin: high-frequency fluctuations (GHz) predominantly limit $T_1$, lower-frequency noise (kHz--MHz) constrains $T_2$, and slowly varying or quasi-static fluctuations determine $T_2^*$, thereby setting the fundamental ODMR linewidth ($\Delta\nu \sim 1/\pi T_2^*$) and sensitivity. In bulk diamond, NV spin dynamics are relatively well understood within a framework dominated by volumetric spin impurities. In NDs, however, nanoscale confinement fundamentally reshapes this noise landscape. As illustrated in Fig.~\ref{Figure: spin impurities and surface noise}a, reducing the particle size, and hence the NV--surface distance $r$, enhances the relative contribution of surface-associated magnetic noise. 
At the same time, distinct non-magnetic noise channels arising from structural disorder (strain) and surface-charge--induced electric-field fluctuations become increasingly important at the nanoscale. These effects establish a competing landscape between spin-impurity-dominated and surface-dominated relaxation mechanisms, leading to NV spin properties in NDs that differ markedly from bulk diamond and vary strongly from particle to particle (Table~\ref{tab:spin_noise_outlook}). In the following, we examine how these noise sources couple to NV spins in NDs, identify the dominant mechanisms governing $T_1$, $T_2^*$, and $T_2$, and discuss material- and surface-level strategies to mitigate their impact and improve sensing performance.

\vspace{-1em}
\subsection*{Magnetic Noise Environment for NV Spins}
\vspace{-1em}

NV spin dynamics are primarily governed by interactions with the surrounding magnetic environment, most notably electronic spins associated with N$_\mathrm{S}^{0}$ (P1 centers) and nuclear spins of \ce{^{13}C}. 
Bulk diamond grown under controlled conditions provides a clean reference for these couplings. In type-IIa bulk diamond with minimal $[\mathrm{N_S}]$ under $^{12}$C-enriched conditions, single NV centers exhibit $T_2$ up to 1.8~ms~\cite{balasubramanian2009ultralong} and $T_2^* \sim 100~\mu$s ($\Delta\nu \sim 3$~kHz), while at natural \ce{^{13}C} abundance (1.1\%), $T_2^*$ is reduced to $\sim 5~\mu$s~\cite{PhysRevB.93.024305}. For NV ensembles, provided that $[\mathrm{NV}] \ll [\mathrm{N_S}]$ to suppress NV--NV dipolar interactions, Bauch \textit{et al.} demonstrated an inverse-linear dependence of $T_2^*$ on the nitrogen impurity concentration for $[\mathrm{N_S}] \gtrsim 0.5$~ppm, with a characteristic scaling of $T_2^* \approx 9.6~\si{\us} \cdot \mathrm{ppm}$ ($\Delta\nu \approx 32$~kHz/ppm)~\cite{bauch2018ultralong,bauch2020decoherence}. The measured dephasing rate exceeds simple pairwise NV--P1 dipolar estimates by a factor of $\sim 1.8$, reflecting the collective dynamics of the P1 spin bath. In the low-nitrogen limit, dephasing is instead dominated by \ce{^{13}C} spins, yielding a scaling of $T_2^* \approx  1~\si{\us} \cdot \mathrm{\%}$ ($\Delta\nu \approx 320$~kHz/\%)~\cite{bauch2018ultralong}. Variations in the effective \ce{^{13}C} hyperfine coupling, spanning tens to hundreds of kHz~\cite{PhysRevB.93.024305,bauch2018ultralong}, reflect the stochastic distribution of nearby \ce{^{13}C} spins. 
A similar inverse-linear dependence of $T_2$ on nitrogen impurities is observed for $[\mathrm{N_S}] \gtrsim 0.5$~ppm, with $T_2 \approx 16\times T_2^*$, saturating at $\sim 700~\si{\us}$ at lower $[\mathrm{N_S}]$ and largely independent of \ce{^{13}C} dilution~\cite{bauch2020decoherence}. In contrast, room-temperature $T_1$ in bulk diamond is largely governed by intrinsic spin--phonon interactions and depends only weakly on $[\mathrm{N_S}]$, increasing from $\sim 3$ to $\sim 6$~ms as $[\mathrm{N_S}]$ decreases from $\approx 60$ to $\approx 0.1$~ppm~\cite{jarmola2012temperature}. Surface-related magnetic noise can, however, strongly reduce $T_1$ for shallow NV centers, with appropriate surface treatments suppressing this contribution by orders of magnitude~\cite{sangtawesin2019origins}. 
These results establish a well-defined bulk-diamond reference for interpreting NV spin relaxation in NDs, as discussed below.

\vspace{-1em}
\subsection*{Volumetric spin impurities}
\vspace{-1em}

In NDs, a comparably systematic picture has not yet emerged. Early single-NV studies in type-Ib NDs reported short coherence times, with $T_2 \approx 1~\si{\us}$ at $[\mathrm{N_S}] \sim 200$~ppm~\cite{tisler2009fluorescence}, increasing to $\sim 5~\si{\us}$ at $[\mathrm{N_S}] \sim 36$~ppm~\cite{knowles2014observing}. 
In this lower-nitrogen regime, $T_2^* \sim 0.5~\si{\us}$ ($\Delta\nu \sim 640$~kHz) can be extended by driving the nitrogen spin bath, to \ce{^{13}C}-limited values of $T_2^* \approx 1.27~\si{\us}$ ($\Delta\nu \sim 250$~kHz), approaching type-IIa bulk-diamond benchmarks~\cite{PhysRevB.93.024305}.  
Further reductions in nitrogen impurities yield substantially longer coherence times, with $T_2 \approx 50~\si{\us}$ (with maxima up to $180~\si{\us}$) at $[\mathrm{N_S}] \sim 0.12$~ppm~\cite{wood2022long}. 
In high-purity, lithographically defined nanocrystals, single NV centers reach $T_2^* \approx 1.8~\si{\us}$ and $T_2 \approx 80~\si{\us}$ at $[\mathrm{N_S}] \sim 0.005$~ppm~\cite{trusheim2014scalable}, improving further to $T_2^* \approx 6.5~\si{\us}$ and $T_2 \sim 100~\si{\us}$ (up to $\sim 360~\si{\us}$) under \ce{^{12}C} enrichment~\cite{andrich2014engineered}. 
Coherence times in these NDs, however, remain well below bulk limits, indicating additional nanoscale noise sources, as discussed below. 
Consistent with this picture, spin locking applied to \ce{^{12}C}-enriched NDs with $[\mathrm{N_S}] = 0.15$~ppm extends coherence from $T_2 \sim 26~\si{\us}$ (spin echo) and $T_2^{\mathrm{DD}} \sim 100~\si{\us}$ (dynamical decoupling, XY8-4) to $T_2^{\mathrm{SL}} \approx 500~\si{\us}$ (spin locking, with maxima approaching $800~\si{\us}$)~\cite{PhysRevApplied.20.044045}, revealing noise components outside conventional echo and dynamical-decoupling bandwidths. These NDs further exhibit $T_1$ values reaching $\sim 3$~ms (with maxima up to $\sim 4.3$~ms), which nevertheless remain shorter than bulk-diamond values~\cite{PhysRevApplied.20.044045,sangtawesin2019origins,rosskopf2014investigation}. 
Altogether, this single-NV picture in NDs underscores the combined effects of volumetric spin impurities and surface-related nanoscale noise sources in limiting NV spin relaxation times.

\begin{table*}[t]
\caption{\textbf{Comparison of NV spin properties and noise amplitudes across bulk diamond and nanodiamond platforms.} Unless otherwise stated, materials contain natural $^{13}$C abundance (1.1\%) and oxygen-terminated surfaces following standard oxidative cleaning protocols. Values in parentheses indicate the maximum reported values across measured particle populations.}
\label{tab:spin_noise_outlook}
% \begin{ruledtabular}
\begin{tabular}{
p{1.4cm}
p{1.2cm}
p{0.8cm}
p{1.2cm}
p{1.2cm}
p{1.0cm}
p{1.0cm}
p{1.0cm}
p{1.0cm}
p{0.8cm}
p{0.8cm}
p{3.6cm}
p{0.8cm}
}
\toprule
\textbf{Material} &
\textbf{Method} &
\textbf{Size} &
\textbf{[NV]} &
\textbf{$[\mathrm{N_S}]$} &
\textbf{$T_1$ } &
\textbf{$T_2^*$} &
\textbf{$T_2$} &
$\boldsymbol{\xi_\perp}$ &
$\boldsymbol{\sigma_{\xi_\perp}}$ &
$\boldsymbol{\sigma_{\beta_z}}$ &
\textbf{Remarks} &
\textbf{Ref.} \\

\textbf{} &
\textbf{} &
\textbf{(nm)} &
\textbf{($\mathrm{ppm})$} &
\textbf{($\mathrm{ppm})$} &
\textbf{(ms)} &
\textbf{($\mu$s)} &
\textbf{($\mu$s)} &
\textbf{(kHz)} &
\textbf{(kHz)} &
\textbf{(kHz)} &
\textbf{} &
\textbf{} \\

\midrule

$^{12}$C diamond &
CVD &
Bulk &
Single &
$<0.001$ &
-- &
$\sim$100 &
-- &
$\sim$50 &
$\sim$8.0 &
$\sim$2.2 &
Electric/strain-noise dominated &
\citenum{PhysRevB.93.024305} \\

\hline

Diamond &
CVD &
Bulk &
Single &
$<0.001$ &
-- &
$\sim$5 &
-- &
$\sim$120 &
$\sim$6.0 &
$\sim$40 &
Magnetic-noise dominated &
\citenum{PhysRevB.93.024305} \\

\hline

ND &
HPHT &
$\sim$30 &
Single &
type-Ib &
-- &
$\sim$0.55 &
-- &
$\sim$7000 &
$\sim$1360 &
$\sim$410 &
Electric/strain-noise dominated &
\citenum{PhysRevB.93.024305} \\

\hline

ND &
HPHT &
$\sim$23 &
Single &
$\sim$36 &
0.15--1.3 &
$\sim$0.44 &
$\approx4$ ($\sim$5.8) &
-- &
-- &
--  &
$T_2^{DD} > 60\,\mu$s $(n_\pi = 71)$ &
\citenum{knowles2014observing} \\

\hline

ND &
CVD &
$\sim$200 &
Single &
$\sim$0.121 &
-- &
-- &
$\approx50$ ($\sim$180) &
-- &
-- &
$\sim$720 &
$T_2^{DD} \sim 460\,\mu$s $(XY8\text{-}4)$ &
\citenum{wood2022long} \\

\hline

$^{12}$C ND &
CVD &
$\sim$100 &
Single &
$\sim$0.15 &
$\approx 2.8$ ($\sim$4.3) &
$\approx 1.35$ ($\sim$2) &
$\approx 26$ ($\sim$40) &
-- &
-- &
--  &
Spin-locked $T_2^{SL} \sim 500 \mu$s &
\citenum{PhysRevApplied.20.044045} \\

\hline

ND &
CVD &
$\sim$50 &
Single &
$<0.005$ &
-- &
$\approx 1.8$ &
$\approx 80$ &
-- &
-- &
-- &
Nanopillars &
\citenum{trusheim2014scalable} \\

\hline

$^{12}$C Diamond &
CVD &
Bulk &
$\sim$0.03 &
$\sim$10&
-- &
0.3--1.2 &
15--18 &
-- &
-- &
-- &
Inverse-linear scaling of $T_2^*$ with $[\mathrm{N_S}]$ and \ce{^{13}C}&
\citenum{bauch2018ultralong} \\

\hline

$^{12}$C Diamond &
CVD &
Bulk &
-- &
$\sim$1 &
-- &
$\sim$9.6 &
$\sim$160 &
-- &
-- &
-- &
$[\ce{NV^-}] \ll [\mathrm{N_S}]$, $T_2 \approx 700 \mu$s for $[\mathrm{N_S}] \sim 0.01~\mathrm{ppm}$  &
\citenum{bauch2020decoherence} \\

\hline

Diamond &
HPHT &
Bulk &
$\approx 0.3$ &
10--30 &
$\sim 5.5$ &
-- &
-- &
-- &
-- &
-- &
-- &
\citenum{jarmola2012temperature} \\

\hline

ND &
CVD &
$\sim$100 &
0.26 &
1.85 &
$\approx 4.7$ &
-- &
-- &
$\sim$3850 &
-- &
-- &
$T_1$ comparable to source bulk diamond&
\citenum{mameli2026properties}  \\

\hline

ND &
HPHT &
$\sim$100 &
$\sim$3 &
$\sim$300 &
$\approx 0.13$ ($\sim$0.3) &
-- &
$\approx 0.28$ ($\sim$0.5) &
$\sim$4400 &
-- &
-- &
Consistent across ND sizes &
\citenum{oshimi2023quantum} \\

\hline

ND &
HPHT &
$\sim$50 &
$\sim$3 &
$\sim$10 &
$\approx 0.17$ &
-- &
$\approx 1.14$ &
$\sim$9100 &
-- &
-- &
-- &
\citenum{bao2025quantumgrade} \\

\hline

ND &
PTQ &
$\sim$50 &
$\sim$0.3 &
$\sim$10 &
$\approx 0.25$ &
-- &
$\approx 1.23$ &
$\sim$7600 &
-- &
-- &
Rapid, large-scale fabrication &
\citenum{bao2025quantumgrade} \\

\hline

$^{12}$C ND &
HPHT &
$\sim$277 &
$\sim$1 &
30--60 &
$\approx 0.68$ ($\sim$1.6) &
$\sim$0.27 &
$\approx 3.2$ ($\sim$5.4) &
$\sim$3100 &
-- &
-- &
$T_2^{DD} > 75\,\mu$s $(n_\pi = 400)$ &
\citenum{oshimi2023quantum} \\

\hline

ND-Silica &
-- &
$\sim$44 &
Ensemble &
$> 100$ &
$\approx 0.38$ &
-- &
$\approx 1.42$ &
-- &
-- &
-- &
$\sim$3$\times$ longer $T_1$ and $T_2^{DD}$ than bare NDs &
\citenum{zvi2025engineering} \\

\hline

ND-Silica &
-- &
$\sim$100 &
Ensemble &
$> 100$ &
$\approx 1$ &
-- &
-- &
-- &
-- &
-- &
$\sim 2 \times$ longer $T_1$ than bare NDs &
\citenum{barzegaramiriolya2025functionalized} \\
\bottomrule
\end{tabular}
\end{table*}

For practical sensing applications, particularly in biological environments, sufficient PL brightness is required, necessitating higher NV densities. Conventional type-Ib NDs with typical $[\ce{NV}] \sim 3$~ppm and $[\mathrm{N_S}] \gtrsim 300$~ppm, however, exhibit very short spin relaxation times, with $T_1 \approx 0.13$~ms and $T_2 \approx 0.3~\si{\us}$, while $T_2^*$ is extremely short ($<100$~ns) and often beyond practical measurement limits~\cite{oshimi2023quantum}(Table~\ref{tab:spin_noise_outlook}). 
Very recently, Oshimi \textit{et al.} demonstrated \ce{^{12}C}-enriched, low-strain NDs with comparatively high $[\mathrm{N_S}] \sim 30$--60~ppm, while maintaining $[\ce{NV}] \approx 1$~ppm and sufficient PL brightness for cellular environments (Fig.~\ref{Figure: spin impurities and surface noise}b)~\cite{oshimi2023quantum}. 
These NDs exhibited $T_2 \approx 3.2~\si{\us}$ (with maxima up to $5.4~\si{\us}$), approaching bulk-limited values at similar $[\mathrm{N_S}]$~\cite{bauch2020decoherence}, while some particles exhibited $T_2^* \sim 170$~ns~\cite{oshimi2023quantum}.
In addition, significantly enhanced $T_1 \approx 0.7$~ms (with maxima up to $1.6$~ms) was observed, although these values remain below bulk-diamond benchmarks~\cite{jarmola2012temperature}, highlighting the persistent influence of surface-related noise in nanoscale environments.
This volumetric spin engineering is also relevant for applications such as nuclear spin hyperpolarization, which exploit spin polarization transfer from NV electron spins to isotopically enriched \ce{^{13}C} nuclear spins~\cite{yuliya2021jpcc,MINDARAVA2020395,doi:10.1126/sciadv.adq6836}.
Beyond nitrogen impurities and the \ce{^{13}C} nuclear spin bath, additional defect-related complexes can further degrade NV performance. Paramagnetic centers such as \ce{NVH^-}, commonly formed during CVD growth, as well as \ce{VH^-} and WAR1 centers, have been shown to couple directly to the NV electron spin~\cite{PhysRevMaterials.8.026203,PhysRevB.103.L100411}. 
Irradiation-induced multivacancy complexes (\ce{V_n}, \ce{V_n^-}) can also perturb the spin and charge dynamics of NV and other color centers (e.g., SnV, GeV)~\cite{pieplow2025quantum,PhysRevB.88.075206,SACHERO2026121054,Shimazaki_2024}. 
The prevalence and relative importance of these defects in NDs, however, remain poorly understood.

\vspace{-1em}
\subsubsection*{Surface-induced noise}
\vspace{-1em}

Paramagnetic surface dipoles, such as dangling bonds with unpaired electrons, couple to NV spins via magnetic dipolar interactions and predominantly affect $T_1$. Using a spherical model with a uniform distribution of surface spins, Tetienne \textit{et al.} showed that surface-induced relaxation of single NV centers in NDs follows a strong size dependence, $T_1^{-1} \propto d_{\rm ND}^{-4}$, reflecting both the distance scaling of dipolar interactions and the three-dimensional proximity of NV centers to the surface (Fig.~\ref{Figure: spin impurities and surface noise}c)~\cite{tetienne2013spin}. Consequently, as particle size is reduced, surface-induced $T_1$ relaxation becomes increasingly dominant, and the NV–surface distance leads to large particle-to-particle variability in $T_1$ even for nominally identical sizes. This model was experimentally validated by introducing controlled paramagnetic noise using \ce{Gd^{3+}} ions, confirming that $T_1$ is governed by the magnetic-noise spectral density at the NV transition frequency. While this model assumes idealized spherical particles, NDs produced by top-down methods often exhibit flake-like or irregular morphologies~\cite{ELDEMRDASH2023268}, which can further broaden the distribution of NV–surface distances.
Accordingly, post-processing surface treatments, particularly oxidative annealing, which improves surface smoothness and suppresses surface-related magnetic noise, have been shown to significantly enhance $T_1$ for shallow NV centers in bulk diamond~\cite{sangtawesin2019origins} as well as in NDs~\cite{alkahtani2025advanced}. Beyond magnetic noise, $T_1$ relaxation of near-surface NV centers is also sensitive to non-magnetic environmental fluctuations. Bulk-diamond studies have shown that $T_1$ of near-surface NV centers is significantly modified by diamagnetic electrolytes, which suppress surface-related noise sources, including electric-field noise~\cite{freire2023role}. Such non-magnetic noise contributions are expected to be more complex in NDs, requiring systematic study for quantitative sensing.

\begin{figure*}[th!]
\centering
\includegraphics[width=\linewidth]{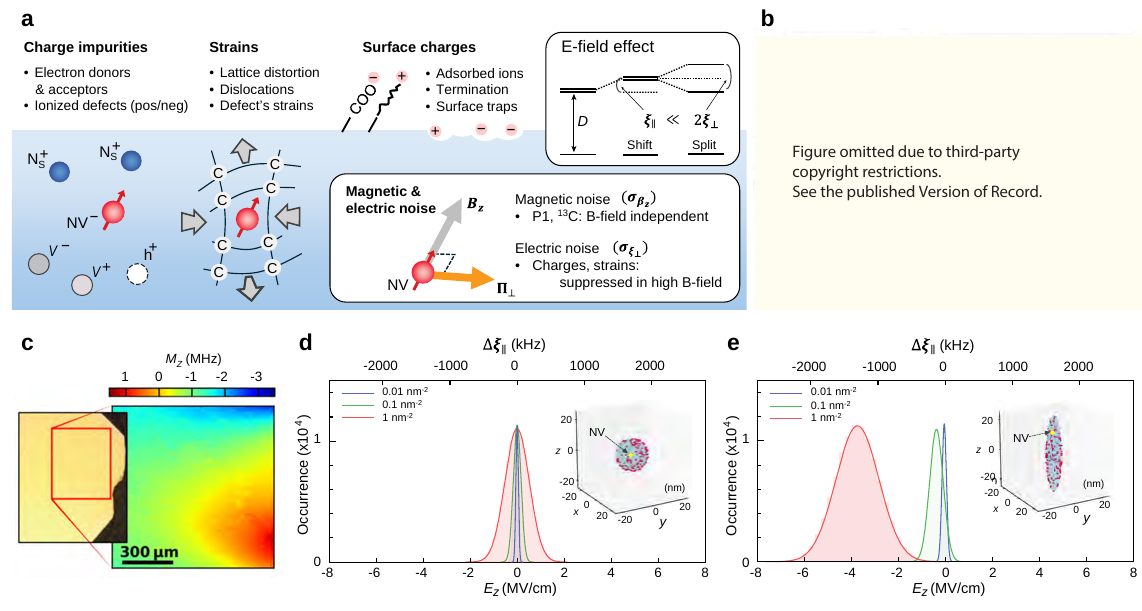}
\vspace{-2em}
\caption{\textbf{Local electric-field and strain coupling to NV spin resonances and their competition with magnetic fields.}
\textbf{a.} Schematics of electric- and strain-field ($\Pi$) coupling to NV spins, 
in competition with an axial magnetic field ($B_z$), 
giving rise to axial and transverse spin-level perturbations, $\xi_\parallel$ and $\xi_\perp$.
\textbf{b.} ODMR spectra and $B_z$ dependence of NV transition frequencies, 
showing suppression of electric- and strain-induced level mixing with increasing $B_z$. 
Adapted from Ref.~\citenum{PhysRevB.93.024305} with permission. Copyrighted by the American Physical Society.
\textbf{c.} Strain imaging over a 1-mm$^2$ area in bulk diamond using ODMR-derived $\xi_\parallel$ (right), 
with the corresponding optical image (left). 
Adapted from Ref.~\citenum{bauch2018ultralong}, licensed under CC BY 4.0.
\textbf{d,e.} Monte Carlo simulations of the axial electric-field component $E_z$ 
for spherical ($r=10$~nm) and ellipsoidal ($a=b=5$~nm, $c=25$~nm) NDs 
with randomly distributed surface charges. 
Insets show NV positions at $z_{\mathrm{NV}}=0$~nm (\textbf{d}) and $z_{\mathrm{NV}}=+12.5$~nm (\textbf{e}). 
Probability distributions of $E_z$ are shown for surface charge densities of 0.01–1~nm$^{-2}$ (solid lines indicate Gaussian fits); top axes indicate the corresponding ODMR frequency shift, 
$\Delta \xi_\parallel$.
}
\label{figure charge-strain}
\end{figure*}

By contrast, transverse coherence $T_2$ is primarily limited by low-frequency surface noise in the kHz--MHz range. Experiments on shallow NV centers in bulk diamond have shown that surface roughness and residual $sp^2$ carbon introduce trapped charges that strongly degrade $T_2$, whereas smooth, well-controlled oxygen-terminated surfaces can substantially improve coherence even for NVs located within a few nanometers of the surface~\cite{sangtawesin2019origins}. 
Complementary theoretical work indicates that both magnetic and electric surface noise contribute to quasi-static dephasing, arising from slowly fluctuating surface spins and charges~\cite{chrostoski2022surface}. 
These surface-related noise mechanisms therefore constitute an intrinsic limitation to NV spin coherence, which is further amplified in NDs by the close proximity of NV centers to the surface. 
This understanding has motivated surface-engineering strategies aimed at mitigating surface-state-related noise. 
For example, graphene capping has been shown to enhance the coherence of shallow NV centers in bulk diamond by suppressing paramagnetic surface spins through interfacial hybridization and charge redistribution~\cite{hao2025coherence}. 
Similarly, in NDs, encapsulation with \ce{SiO_2} shells (Fig.~\ref{Figure: spin impurities and surface noise}d) has been shown to reduce surface spin noise via interfacial band bending, leading to $\sim$3$\times$ enhancements in coherence under dynamical decoupling ($T_2^{\mathrm{DD}}$)~\cite{zvi2025engineering} as well as in $T_1$~\cite{zvi2025engineering,barzegaramiriolya2025functionalized}. 
In contrast, $T_2$ is only weakly affected by \ce{SiO_2} encapsulation~\cite{zvi2025engineering}, suggesting that low-frequency noise components ($<3.7$~MHz) remain largely unchanged upon coating. 
Furthermore, double-quantum relaxation ($T_1^{\mathrm{DQ}}$), which is predominantly sensitive to electric-field noise, shows no appreciable change upon \ce{SiO_2} encapsulation~\cite{zvi2025engineering}, in agreement with bulk-diamond studies reporting residual electric-field fluctuations 
despite well-controlled oxygen-terminated surfaces~\cite{sangtawesin2019origins}.
This behavior underscores the challenge of fully mitigating surface-induced decoherence, particularly electric-field noise, which is strongly amplified at the nanoscale, as discussed below.

\vspace{-1em}
\subsection*{Local Electric Fields and Strain Environments for NV Spins}
\vspace{-1em}

Beyond the magnetic environment described above, NV spins also interact with electric-field ($\mathcal{E}_i$) and strain ($M_i$) perturbations, which couple through the same linear Stark interaction and enter as effective fields $\Pi_i$ in the NV spin Hamiltonian~\cite{PhysRevB.93.024305,dolde2011electric} (Eq.~\ref{eq:hamitonian}). The NV center is far more sensitive to transverse than to axial fields by a factor of $\sim 50$~\cite{van1990electric}: axial components primarily shift the ODMR resonance frequency ($\xi_\parallel$), whereas transverse components induce a level splitting of $\ket{m_s=\pm1}$ by $2\xi_\perp$ (Fig.~\ref{figure charge-strain}a). While this coupling enables sensing of external electric and strain signals~\cite{dolde2011electric}, it also renders NV spins sensitive to internal material disorder. These non-magnetic perturbations become increasingly important once magnetic noise is suppressed, particularly in NDs.

\vspace{-1em}
\subsubsection*{Volumetric charge and strain environments}
\vspace{-1em}

Materials fabrication inevitably introduces lattice imperfections, giving rise to strain fields within the diamond lattice. In parallel, stabilization of the NV$^{-}$ charge state occurs through donor--acceptor charge transfer (Eq.~\ref{eq:N-NV-charge exchange}), most commonly involving neutral N$_\mathrm{S}$ (N$_\mathrm{S}^{0}$), which becomes ionized (N$_\mathrm{S}^{+}$) and generates local electric field set by growth conditions and defect engineering.
For example, in high-purity bulk diamond with minimal $[\mathrm{N_S}]$, single NV centers experience relatively weak transverse perturbations ($\xi_\perp \sim 100$~kHz), which can be readily suppressed by applying weak magnetic fields ($\sim$0.01~mT) (Fig.~\ref{figure charge-strain}b). By contrast, single NV centers in type-Ib NDs exhibit strongly amplified disorder ($\xi_\perp \sim$ 7 MHz), requiring Magnetic fields of several~mT to suppress these effects~\cite{PhysRevB.93.024305}. 

Bulk diamond with controlled growth and nitrogen incorporation enables clearer separation of electric-field and strain-induced contributions in NV ensemble ODMR spectra. Mittiga \textit{et al.} showed that zero-field ODMR spectra of NV ensembles, which had often been attributed to lattice strain, are quantitatively captured by a microscopic charge model~\cite{PhysRevLett.121.246402}. 
The extracted charge densities were comparable to the NV concentration, supporting a charge-transfer picture (Eq.~\ref{eq:N-NV-charge exchange}).
A recent study further demonstrated the cumulative effect of microscopic electric fields as a function of $[\mathrm{N_S}]$,with [NV] fixed at $\sim$3~ppm~\cite{yu2024optically}. 
Furthermore, in systems with high $[\mathrm{N_S}] \sim 200$~ppm, ODMR spectra exhibit non-Gaussian skirts and dip splitting together with optical-power-dependent spectral broadening and splittings, e.g. $\xi_\perp \approx 5$--8~MHz in NDs~\cite{yu2024optically}.
These power-dependent spectral features are attributed to laser-induced neutralization of N--V$^{-}$--N$^{+}$ pairs into N--V$^{0}$--N$^{0}$ pairs, which leads to a dynamically evolving internal electric-field environment~\cite{fujiwara2020real,ito2023optical,yu2024optically}. 
By contrast, in bulk diamond with lower $[\mathrm{N_S}] \sim 14$~ppm and comparable NV density, ODMR spectra show discrete peak splitting ($\xi_\perp \approx 2.5$~MHz) with weak power dependence, consistent with reduced electric-field heterogeneity.
Beyond ionized nitrogen donors, other charged defects may also contribute to local electric-field fluctuations, including donor or acceptor impurities (e.g., P$^+$, B$^-$), vacancy-related complexes (V$^\pm$, V$_2^\pm$), and mobile carriers. 

The role of strain in limiting NV spin coherence is well established in bulk diamond, whereas its relative impact in NDs has yet to be quantitatively clarified.
In bulk diamond, strain gradients can be mapped directly using NV spins as local probes~\cite{bauch2018ultralong,Kehayias2019} (Fig.~\ref{figure charge-strain}c). These measurements show that once magnetic interactions are suppressed, spatial strain gradients become a dominant source limiting $T_2^*$, particularly as $[\mathrm{N_S}]$ falls below $\sim$1~ppm~\cite{bauch2018ultralong,Shinei2025}. 
Through advanced growth engineering, NV-ensemble bulk diamonds have consequently achieved $T_2^*$ approaching the fundamental \ce{^{13}C}-limited regime, emphasizing the critical role of structural perfection~\cite{blinder2024reducing}. 
Analogous improvements are now emerging in NDs. With reduced strain/electric field disorder, either grown via unconventional bottom-up approaches~\cite{alkahtani2019growth} or produced by direct milling of \ce{^{12}C}-enriched NV-ensemble diamond crystals~\cite{oshimi2023quantum}, exhibit markedly improved NV spin coherence and control  compared to commonly produced type-Ib NDs. The relative impact of strain gradients and electric field disorder in NDs, however, remains to be quantitatively assessed.

\vspace{-1em}
\subsubsection*{Surface-charge effects}
\vspace{-1em}

At the nanoscale, surface charge environments play a role comparable to volumetric charges by coupling to NV spins via $\xi_\perp$, and are therefore particularly relevant in NDs with near-surface NV centers. Recent studies have shown that interactions between ND surfaces and biomolecular environments can induce measurable changes in ODMR spectra~\cite{millikervin2025Sow,zvi2026probing}, arising from surface charge–induced electric-field variations, and potentially appearing as artifacts in thermometry in biological environments. Surface capping strategies, including uniform molecular functionalization~\cite{millikervin2025Sow} and sol--gel silica encapsulation~\cite{zvi2026probing}, have been shown to mitigate such electric-field perturbations. Surface-charge management is therefore critical to reliable ND-based quantum sensing.

To assess surface charge effects in NDs, we performed simple numerical simulations of internal electric fields. In an idealized spherical ND with a uniform surface charge distribution, Gauss’s law predicts a zero internal electric field. In realistic NDs, however, deviations from this ideal case--including nonspherical geometries, surface roughness, and inhomogeneous surface charge distributions--generate non-zero internal electric fields. Following Ref.~\citenum{PhysRevLett.121.246402}, we modeled spherical and ellipsoidal NDs with randomly distributed surface charges at average areal densities of 0.01–1 nm$^{-2}$, corresponding to approximately 10--1000 elementary charges per particle for NDs with a radius of 10 nm.
For both geometries, configurations with the NV center located at the center and at off-axis positions along the symmetry axis were evaluated.
Increasing surface charge density broadens the distribution of the axial electric-field component $E_z$ in all cases, including spherical NDs with the NV center located at the center (Fig.~\ref{figure charge-strain}d).
In off-axis configurations, the axial electric-field distribution is shifted, with this effect being particularly pronounced for ellipsoidal NDs (Fig.~\ref{figure charge-strain}e).
In the single-NV limit, additional spectral asymmetries can arise from the relative orientation between the NV axis and the microwave polarization~\cite{PhysRevLett.121.246402,millikervin2025Sow}, but these effects are expected to be reduced by ensemble averaging.
Although not explicitly included in the model, the simulations capture the dominant influence of surface-charge-induced field inhomogeneity and bias fields in NDs, underscoring the importance of surface engineering toward quantum-grade performance.

\vspace{-1em}
\section*{Outlook and Concluding Remarks}
\vspace{-1em}

Quantum sensing with solid-state spin defects, led by NV centers, has matured over the past two decades into a powerful nanoscale sensing platform, with sensitivities now reaching the single-molecule regime~\cite{wolfowicz2021quantum,du2024single}. 
While bulk diamond enabled foundational advances and benchmark demonstrations, NDs, historically developed for biological applications~\cite{mochalin2020properties}, extend NV-based sensing to mobile probes capable of operating in complex and realistic environments~\cite{feng2022recent,yingke2022advscireview}. 
Under such conditions, fundamental constraints, including ND rotational diffusion and the absence of reliable magnetic-field alignment, limit coherence-based protocols and restrict the dominant sensing modalities to cw-ODMR and all-optical $T_1$ relaxometry. In biological media, strong background autofluorescence further necessitates high PL brightness, favoring NDs hosting NV ensembles over single defects. 
Within these constraints, ND-based quantum sensing has enabled impactful applications, particularly subcellular nanothermometry in living systems using zero-field cw-ODMR, with demonstrations spanning metabolic thermogenesis~\cite{lee2025organelle}, therapeutic monitoring~\cite{wu2021nanodiamond}, and related processes~\cite{fujiwara2021diamond}. 
At the same time, relative to bulk diamond, strong spectral broadening and reduced ODMR contrast in NDs, arising from enhanced lattice strain and degraded NV spin properties, substantially reduce sensing sensitivity~\cite{fujiwara2021diamond}. 
Recently, Sow \emph{et al.} demonstrated that, in cellular environments, apparent ODMR spectral shifts can also originate from surface electric-field fluctuations in addition to temperature changes, underscoring the importance of disentangling non-thermal contributions for reliable readout~\cite{millikervin2025Sow}. 
In parallel, $T_1$ relaxometry with NDs has emerged as a powerful platform for detecting paramagnetic species~\cite{rendler2017optical}, including free radicals, ions, and specific molecules, and for monitoring associated processes such as cellular metabolic activity~\cite{nie2021quantum}, redox chemistry~\cite{barton2020nanoscale}, and medical diagnostics~\cite{zalieckas2024quantum}, as reviewed elsewhere~\cite{mzyk2022relaxometry}. 
Recent studies further highlight that $T_1$ is also sensitive to the local chemical environment, including electrolyte composition~\cite{freire2023role} and pH-dependent surface charge variations~\cite{fujisaku2019ph}, as well as optically induced charge-state effects under varying excitation conditions~\cite{PhysRevB.108.075411}. Pronounced particle-to-particle variability of $T_1$, however, underscores the importance of materials control for reliable quantitative sensing and reproducibility across experiments~\cite{mzyk2022relaxometry}.

Importantly, several effects traditionally viewed as limitations have increasingly been repurposed as sensing resources. NV charge-state dynamics, when appropriately controlled, enable simplified all-optical, PL-based sensing of electrochemical potentials~\cite{karaveli2016modulation} and pH values~\cite{Sow-nanoscale-2020} using NDs with single NV centers. Likewise, Brownian motion of NDs, while limiting coherence-based protocols, has been exploited for nanoscale rheological analysis via single-particle tracking combined with thermometry~\cite{gu2023simultaneous}. Additionally, cw-ODMR in applied magnetic fields has been used to probe ND rotational dynamics, providing a powerful platform for studying cellular mechanics~\cite{Cui2022NanoLett,cui2025nanodiamond}. In carefully controlled environments, magnetic-field alignment enables coherence-based protocols such as NV-detected NMR with NDs~\cite{holzgrafe2020nanoscale}, including demonstrations of multi-species detection within volumes approaching $(20~\mathrm{nm})^3$. Emerging approaches that integrate microfluidics~\cite{fujiwara2023diamond} and machine learning~\cite{yamamoto2025nanodiamond} offer additional routes to enhance sensitivity, robustness, and data interpretation under controlled conditions.

Together, these advances underscore that progress in ND-based quantum sensing in complex environments, beyond proof-of-concept demonstrations, depends not only on protocol-level innovation and instrumental integration, but ultimately on materials-level control. Sensing performance and reliability are fundamentally constrained by NDs materials properties. Fully unlocking their potential requires NDs with predictable and stable NV charge states, application-optimized spin properties, and minimal particle-to-particle variability. Achieving this, in turn, demands \textit{controlled and reproducible ND synthesis at scale}, encompassing NV formation with minimal collateral damage, together with surface treatments that provide electronic passivation while enabling functional specificity for operation in complex chemical and biological environments.

Material engineering of NDs for quantum sensing can be supported by improved control over ND production and, at the same time, guided by a mechanism-level understanding of the noise processes governing NV spin properties, which are comparatively well established in bulk diamond through controlled growth and systematic experiments~\cite{barry2020sensitivity}. Naturally, magnetic noise arising from spin impurities (P1 centers, \ce{^{13}C}, and paramagnetic surface dipoles) drives NV spin dynamics; in parallel, proximity charge-induced electric fields and strain-equivalent perturbations couple to the NV spin through the transverse zero-field splitting term ($\xi_\perp$). These two contributions, commonly described in terms of magnetic noise ($\sigma_{\beta_z}$) and effective electric disorder ($\sigma_{\xi_\perp}$), have been quantitatively established, most clearly through their competing influence on $T_2^*$~\cite{PhysRevB.93.024305}, with representative regimes summarized in Table~\ref{tab:spin_noise_outlook}.

In ultrapure, \ce{^{12}C}-enriched diamond with negligible $[\mathrm{N_S}]$ and modest static strain ($\xi_\perp \sim 100$~kHz), NV dephasing is dominated by electric noise ($\sigma_{\xi_\perp}\sim8$~kHz), yielding $T_2^* \sim28~\si{\us}$. Applying weak magnetic fields ($\sim0.01$~mT) suppresses this contribution and extends $T_2^*$ to $\sim100~\si{\us}$, corresponding to $\sigma_{\beta_z}\sim2.2$~kHz. Introducing natural \ce{^{13}C} abundance (1.1\%) reverses this balance, rendering magnetic noise dominant and reducing $T_2^*$ to $\sim5~\si{\us}$ ($\sigma_{\beta_z}\sim40$~kHz). Systematic bulk-diamond studies further establish near-inverse scaling of $T_2^*$ and $T_2$ with $[\mathrm{N_S}]$, providing a well-defined spin-impurity-dominated framework for coherence optimization~\cite{bauch2020decoherence}. At the nanoscale, this clean separation of noise mechanisms breaks down. 
For single NV centers in typical type-Ib NDs, magnetic noise amplitudes increase by roughly an order of magnitude ($\sigma_{\beta_z}\sim400$~kHz), while enhanced lattice strain and surface-charge fluctuations generate effective transverse fields $\xi_\perp$ in the MHz range~\cite{PhysRevB.93.024305}. 
As a result, NV dephasing is often dominated by electric noise ($\sigma_{\xi_\perp}\sim1360$~kHz). Magnetic fields of several~mT, in contrast to $\sim$0.01~mT in bulk diamond, are required to partially suppress it, yielding $T_2^* \lesssim 1~\si{\us}$. 
Even in NDs with low nitrogen concentrations ($[\mathrm{N_S}] \sim 0.15$ to $<0.005$~ppm), single-NV coherence times remain limited to $T_2 \approx 25$--80~\si{\us}, with maxima up to $\sim180~\si{\us}$ (Table~\ref{tab:spin_noise_outlook}). 
These values are well below the millisecond-scale coherence achievable in high-purity bulk diamonds~\cite{balasubramanian2009ultralong,bauch2020decoherence}.
In practical biosensing applications, higher NV densities are typically required to ensure sufficient PL brightness, introducing additional NV--NV dipolar interactions that further complicate optimization. This competition between magnetic noise and electric or strain-induced disorder, and the difficulty of disentangling their respective contributions, is therefore a defining feature of ND-based quantum sensing. These challenges are exacerbated in NDs produced by dominant top-down synthesis routes, which introduce irregular morphologies, heterogeneous surface states, and enhanced structural disorder~\cite{https://doi.org/10.1002/ppsc.201900009}, all of which are difficult to capture within simplified theoretical models~\cite{tetienne2013spin}. Conventional defect engineering based on irradiation further complicates this landscape, as lattice damage and compensating defects emerge at substantially lower fluences than in bulk diamond. Achieving reproducible and quantitative ND-based sensing thus requires coordinated materials optimization across multiple length scales, encompassing control of volumetric impurity and defect populations, minimization of strain and morphological disorder, and deliberate surface engineering that stabilizes the NV charge state while managing interactions with complex chemical and biological environments.

One strategy to simplify process control is to create NV centers prior to milling, avoiding the pronounced effect of irradiation in NDs while also enabling improved control over volumetric spin impurities, albeit often at the cost of reduced NV density after size reduction~\cite{shenderova2019synthesis}. Using this approach, low-strain, \ce{^{12}C}-enriched NDs have been realized~\cite{oshimi2023quantum} with sufficient PL brightness ($[\ce{NV}] \approx 1$~ppm) for cellular environments while approaching bulk-limited spin lifetimes, with $T_1$ and $T_2$ enhanced by factors of at least $\sim$5$\times$ and $\sim$10$\times$, respectively, compared with commercial type-Ib NDs. Notably, comparable ODMR contrast can be achieved with approximately 20$\times$ lower microwave power, reducing external heating and benefiting thermometry applications. However, $T_1 \approx$ 0.68~ms (with maxima up to $\sim$1.6~ms) remains limited, compared with $\approx$ 3--5~ms in bulk diamond with similar $[\mathrm{N_S}]$ or $[\mathrm{NV}]$ concentrations. Recent work using a similar approach demonstrates that milled NDs can retain average $[\mathrm{N_S}] \approx 2$~ppm and $[\ce{NV}] \approx 0.25$~ppm, and more importantly, $T_1 \sim 4.7$~ms, comparable to the source bulk diamond~\cite{mameli2026properties}. However, challenges remain in stabilizing the NV$^{-}$ charge state in NDs with low $[\mathrm{N_S}]$ and in achieving consistent ODMR response across particles. Overall, these studies reflect the critical role of surface engineering in NDs in shaping NV spin and charge properties. These surface-engineering approaches range from standard contamination-removal protocols~\cite{alkahtani2025advanced} to advanced encapsulation strategies such as \ce{SiO_2} coating~\cite{zvi2025engineering}, which define surface termination and modify surface electronic states through band bending. Such encapsulation suppresses high-frequency noise, yielding $\sim$3$\times$ enhancements in $T_1$ and extended coherence under dynamical decoupling. By contrast, low-frequency noise components that limit spin-echo $T_2$ and inhomogeneous dephasing $T_2^*$ remain largely unaffected, while encapsulation introduces an intrinsic trade-off by increasing the NV–target distance, thereby reducing sensing sensitivity in practice.

Furthermore, conventional defect engineering in NDs has been revisited through thermal processing at unconventionally high temperatures ($\gtrsim1400$--1700~\si{\degreeCelsius}). Such treatments have been shown to partially recover irradiation-induced lattice damage, improving NV spin properties in NDs and micron-sized particles~\cite{gierth2020enhanced,torelli2020high}. In bulk diamond, comparable high-temperature treatments have further demonstrated vacancy generation and NV formation without requiring irradiation~\cite{wong2022microscopic}. 
More recently, a single-step, irradiation-free approach based on plastic deformation of type-Ib NDs under extreme pressures ($\sim$7~GPa) and temperatures ($\sim$1700~\si{\degreeCelsius}) enables rapid, large-scale ND fabrication with improved NV charge stability and spin properties~\cite{bao2025quantumgrade}. 
However, such high-temperature processing promotes vacancy-assisted defect reconfiguration, leading to the formation of complex defects such as NVN (H3)~\cite{dei2019fancy,gierth2020enhanced} and can reduce NV density by up to an order of magnitude, thereby diminishing PL brightness~\cite{bao2025quantumgrade}. At the same time, optically active H3 centers enable dual biolabelling in cellular environments~\cite{bao2025quantumgrade}.  
In parallel to bulk-diamond studies, enhanced NV$^{-}$ formation has been achieved through simultaneous electron irradiation and annealing~\cite{capelli2019increased}, as well as through charge-assisted defect engineering via deliberate incorporation of donor impurities such as phosphorus, oxygen, or sulfur~\cite{herbschleb2019ultra,luhmann2021charge}. Related charge-defect complex dynamics have also been resolved with other color centers, such as tin-vacancy (SnV) in diamond, where time-resolved detection of electric
fields has revealed lattice-scale charge trapping, charge-transport dynamics, and associated noise generation, underscoring the broader relevance of charge-defect engineering across diamond-hosted spin defects~\cite{pieplow2025quantum}. Together, these results suggest additional pathways for controlling defect charge state and density that may inform future ND engineering strategies.

More broadly, pronounced variations in NV properties persist across NDs and remain an open challenge, largely reflecting irregular morphologies inherent to top-down synthesis routes. While recent studies have established correlations between PL brightness and particle shape~\cite{nano11102706,haotin2023acsnano}, the corresponding impact on NV spin properties—and its distinction from surface chemistry and charge effects—remains largely unexplored.

Historically, nanoparticles have been produced using bottom-up synthetic routes that afford comparatively strong control over particle crystallinity and morphology, enabling systematic structure--property relationships. In this context, bottom-up approaches to ND synthesis offer an attractive alternative to top-down processing by avoiding extreme mechanical damage. Promising demonstrations include HPHT conversion of molecular precursors~\cite{https://doi.org/10.1002/adma.201502672,https://doi.org/10.1002/cnma.201700349}, CVD growth~\cite{https://doi.org/10.1002/admi.201901408}, and laser ablation~\cite{Zousman2014PureNanodiamonds}. Interestingly, slow HPHT growth from azaadamantane at unusually low temperatures ($\sim$400~\si{\degreeCelsius}) yields low-strain NDs with narrow NV resonance lines~\cite{alkahtani2019growth}. 
More recently, HPHT conversion of nanographene has enabled large-scale production with improved size uniformity and in situ incorporation of alternative color centers such as GeV and SnV~\cite{liang2026bottom}. In a conceptually distinct approach, hydrothermal growth under mild temperature ($\sim$220~\si{\degreeCelsius}) and pressure ($\sim$2.5~MPa) conditions has demonstrated solution-processed ND formation with NV and SiV centers created in situ~\cite{alkahtanihydrothermal}, indicating that local reaction dynamics can govern diamond nucleation without extreme thermodynamic conditions. These recent results demonstrate the promise of bottom-up ND synthesis, offering improved morphological control and more spherical-like particle geometries, while still predominantly yielding small particles with relatively few color centers, thereby motivating further efforts to scale particle size and NV density to levels compatible with ensemble-based and biological quantum sensing.

In conclusion, the performance of ND-based quantum sensors is governed by a complex interplay between magnetic, electric, and strain-related noise sources that are strongly amplified at the nanoscale by surface proximity. Unlike bulk diamond, where dominant decoherence mechanisms can often be isolated and systematically engineered, NDs operate in a regime where volumetric disorder, surface states, morphology, and environmental coupling are intricately intertwined.
As a result, sensing performance is fundamentally materials-limited rather than protocol-limited, and progress hinges on transforming these nanoscale constraints into design variables. Achieving this will require NDs with predictable and stable NV charge states, optimized and application-specific spin properties, controlled morphology, and minimal particle-to-particle variability, produced through scalable and reproducible synthesis and surface engineering. Continued advances along these directions promise to move ND quantum sensing from proof-of-principle demonstrations toward robust, interpretable, and deployable platforms for biological and nanoscale science, while opening opportunities to exploit nanoscale charge, strain, and dynamics as sensing modalities in their own right.

\vspace{-1em}
\section*{Acknowledgements}
\vspace{-1em}

This work was supported in part by JST-ASPIRE (JPMJAP2339), JSPS KAKENHI (JP20KK0317, JP20H00335, JP24H00406, JP25K00934, JP25K17361), JST-CREST (JPMJCR23I2), Australian Research Council (DP230101847, DP260102670), New Challenge Research (JSR–UTokyo Collaboration Hub, CURIE), the Foundation of Kinoshita Memorial Enterprise, and the Asahi Glass Foundation.

\section*{Author Contributions}
\vspace{-1em}

A.R., K.O., and M.F. contributed to conceptualization of the review.
K.S. and M.F. performed electric noise simulations and analysis.
A.R., K.O., K.S., S.L.Y.C., and M.F. contributed to drafting the manuscript.
All authors participated in discussion and revision of the manuscript and approved the final version.

% \nocite{*}
\bibliography{aipsamp}% Produces the bibliography via BibTeX.

\end{document}